\documentclass[cernpreprint, BSM, texlive=2017, USenglish, texmf]{atlasdoc}
 
\usepackage{atlaspackage}
\usepackage[backref=false,block=space]{atlasbiblatex}
 
\usepackage{atlasphysics}
 
\graphicspath{{../logos/}{../figures/}{figures/}{logos/}}
 
\usepackage{ANA-SUSY-2018-40-PAPER-defs}

\AtlasTitle{
Search for bottom-squark pair production in $p\!p$~collision events at $\sqrt{s} = 13$~TeV~with hadronically decaying $\tau$-leptons,
$b$-jets and missing transverse momentum using the ATLAS detector
}

\journal{\PRD}

\AtlasAbstract{
A search for pair production of bottom squarks in events with hadronically decaying $\tau$-leptons, $b$-tagged jets and large missing transverse momentum is presented.
The analyzed dataset is based on proton--proton collisions at $\sqrt{s}=13$~\tev\ delivered by the Large Hadron Collider and recorded by the ATLAS detector from 2015 to 2018, and corresponds to an integrated luminosity of 139~fb$^{-1}$.
The observed data are compatible with the expected Standard Model background.
Results are interpreted in a simplified model
where each bottom squark is assumed to decay into the second-lightest neutralino $\tilde \chi_2^0$ and a bottom quark,
with $\tilde \chi_2^0$ decaying into a Higgs boson and the lightest neutralino $\tilde \chi_1^0$.
The search focuses
on final states where at least one Higgs boson decays into a pair of hadronically decaying $\tau$-leptons.
This allows the acceptance and thus the sensitivity to be significantly improved relative to the previous results
at low masses of the $\tilde \chi_2^0$,
where bottom-squark masses up to 850~\gev\ are excluded at the 95\% confidence level,
assuming a mass difference of 130~\gev\ between $\tilde \chi_2^0$ and $\tilde \chi_1^0$.
Model-independent upper limits are also set on the cross section of processes beyond the Standard Model.
}
 
\author{The ATLAS Collaboration}

\AtlasRefCode{SUSY-2018-40}
 
\AtlasJournalRef{Phys. Rev. D 104 (2021) 032014}
\AtlasDOI{10.1103/PhysRevD.104.032014}

\PreprintIdNumber{CERN-EP-2020-235}
 
\hypersetup{pdftitle={ATLAS document},pdfauthor={The ATLAS Collaboration}}
 
\begin{document}
 
\maketitle

\section{Introduction}

Although the Standard Model (\SM) of particle physics is a very successful theory,
it does not provide a natural explanation for the large hierarchy between the energy scale of electroweak interactions
and the Planck scale related to the gravitational interaction.
Nor does it have a viable candidate particle for dark matter,
and it does not include a quantum description of gravity.
Supersymmetry (SUSY)~\cite{Golfand:1971iw2,Volkov:1973ix,Wess:1974tw,Wess:1974jb,Ferrara:1974pu,Salam:1974ig}
is a theoretical framework that extends the \SM by introducing partner states for the known particles,
where the partners have the same quantum numbers as the respective \SM particles but differ in spin by half a unit.
This leads to new loop corrections to the Higgs boson mass that cancel out those involving SM particles,
thereby solving the hierarchy problem~\cite{Sakai:1981gr_FIXED,Dimopoulos:1981yj,Ibanez:1981yh,Dimopoulos:1981zb}.
When conservation of $R$-parity~\cite{Farrar:1978xj} is assumed,
the lightest supersymmetric particle is stable
and would be a viable candidate for dark matter if it is weakly interacting~\cite{Goldberg:1983nd,Ellis:1983ew}.
However, SUSY must be a broken symmetry
in order to allow the supersymmetric particles to be heavier than their \SM partners and evade detection so far.
Naturalness arguments~\cite{Barbieri:1987fn,deCarlos:1993yy} support the assumption
that the partner states of the third-generation quarks,
the top squarks \stop and the bottom squarks \sbottom,
should be light and thus have relatively large production cross sections.
They might even be the only strongly produced
supersymmetric states within the current mass reach of the LHC.
 
This \article presents a search for pair production of bottom squarks \sbottom
that decay via the second-lightest neutralino \ninotwo to the lightest neutralino \ninoone.
The neutralinos $\tilde\chi_{1,2,3,4}^{0}$ together with the charginos $\tilde\chi_{1,2}^\pm$
are mixtures of the partner states of the electroweak gauge bosons (bino and winos) and Higgs bosons (higgsinos).
The simplified model~\cite{Alwall:2008ve,Alwall:2008ag,Alves:2011wf} of production and decay of supersymmetric particles
considered in this search is shown in~\cref{fig:sbottom_diagram}.
It is inspired by the Minimal Supersymmetric Standard Model~\cite{Fayet:1976et,Fayet:1977yc}
in scenarios where the branching ratio ${\cal B}(\ninotwo\to h\ninoone)$ is enhanced,
\eg when the \ninoone is bino-like and the \ninotwo a wino--higgsino mixture.
The branching ratio ${\cal B}(\sbottom \to b\ninotwo)$ is large
compared to that of the direct decay ${\cal B}(\sbottom \to b\ninoone)$,
which is studied elsewhere~\cite{SUSY-2016-28},
when the mixture of the bottom squark is such that it is mostly the superpartner of the left-chiral bottom quark,
the \ninoone is mostly bino,
and the \ninotwo mostly wino.
A wino- or higgsino-like \ninotwo will be accompanied by a \chinoonepm,
which allows the decay $\sbottom \to t \chinoonem$.
This decay mode is relevant if the mass difference between the bottom squark and the chargino is larger than the top-quark mass.
In the simplified model, ${\cal B}(\sbottom\to b\ninotwo)$ and ${\cal B}(\ninotwo\to h\ninoone)$
are assumed to be \percent{100}.
Moreover, the Higgs boson is assumed to have the same properties as in the SM, namely
$m(h) = \GeVx{125}$,
${\cal B}(h\to\bbbar)       = \percent{58} $, and
${\cal B}(h\to\tau^+\tau^-) = \percent{6.3}$.
Only decays of the Higgs bosons into $\bbbar$, $\tau^+\tau^-$, $W^+W^-$ and $ZZ$ are considered in the signal-model generation.
Furthermore, the mass difference $\Delta m(\ninotwo, \ninoone)$ between the \ninotwo and \ninoone is set to \GeVx{130}
such that the Higgs boson produced in the decay of the \ninotwo is on its mass shell.
The free parameters of the model are chosen to be the masses \msb and \mneuttwo.

The signal model illustrated in~\cref{fig:sbottom_diagram} yields a final state with two bottom quarks,
two Higgs bosons, and missing transverse momentum from the two \ninoone particles that escape the detector without interacting.
This analysis selects a final state with a pair of $\tau$-leptons arising from the decay of one of the Higgs bosons,
such that it complements a previous ATLAS search~\cite{SUSY-2018-31},
which focuses on final states with multiple \bjets.
This particular decay mode of the Higgs boson has never been exploited by a bottom-squark search until now.
The neutrinos from the $\tau$-lepton decays
provide a source of missing transverse momentum
in addition to the pair of \ninoone.
This increases the acceptance of the search
in the region of parameter space where the \ninotwo is relatively light and the \ninoone moderately boosted,
where the previous ATLAS analysis has limited sensitivity.
The same simplified model has been employed by the CMS Collaboration
in a search targeting $h\to\gamma\gamma$ decays~\cite{CMS-SUS-18-007}.
Using a dataset of \ifbx{77.5}, the CMS analysis excludes bottom-squark masses up to \GeVx{530}
for an almost massless \ninoone at the \percent{95} confidence level,
and bottom-squark masses up to at least \GeVx{400} for heavier masses of the \ninoone.

\begin{figure}[t]
\centering
\includegraphics[width=5cm]{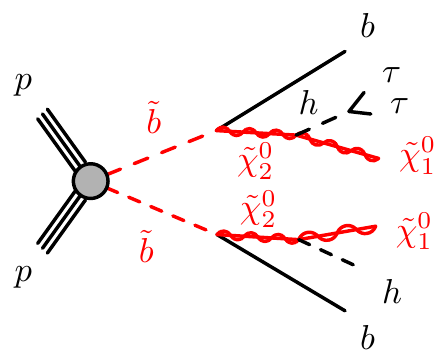}
\caption{Simplified model of bottom-squark pair production and the decay chain targeted by this analysis.}
\label{fig:sbottom_diagram}
\end{figure}

The \article is structured as follows.
After this introduction, Section~\ref{sec:detector} briefly describes the ATLAS detector,
and Section~\ref{sec:samples} presents the dataset and simulated event samples.
The reconstruction of physics objects is described in Section~\ref{sec:reconstruction},
and the signal selection and analysis discriminants are detailed in Section~\ref{sec:selection}.
The procedures to derive the background estimate are explained in Section~\ref{sec:background},
followed by a summary of the systematic uncertainties in Section~\ref{sec:systematics}.
Section~\ref{sec:result} presents the results from the analysis and their interpretation,
and conclusions are given in Section~\ref{sec:conclusion}.

 
\section{ATLAS detector}
\label{sec:detector}
 
The ATLAS experiment~\cite{PERF-2007-01,ATLAS-TDR-2010-19,ATLAS-TDR-2010-19-add,PIX-2018-001} at the LHC is a multipurpose particle detector
with a forward--backward symmetric cylindrical geometry and nearly \(4\pi\) coverage in
solid angle.\footnote{ATLAS uses a right-handed coordinate system with its origin at the nominal interaction point (IP)
in the center of the detector and the \(z\)-axis along the beam pipe.
The \(x\)-axis points from the IP to the center of the LHC ring,
and the \(y\)-axis points upwards.
Cylindrical coordinates \((r,\phi)\) are used in the transverse plane,
\(\phi\) being the azimuthal angle around the \(z\)-axis.
The pseudorapidity is defined in terms of the polar angle \(\theta\) as \(\eta = -\ln \tan(\theta/2)\).
Angular distance is measured in units of \(\Delta R \equiv \sqrt{(\Delta\eta)^{2} + (\Delta\phi)^{2}}\).
}
It consists of an inner tracking detector surrounded by a thin superconducting solenoid
providing a \SI{2}{\tesla} axial magnetic field, electromagnetic and hadronic calorimeters, and a muon spectrometer.
The inner tracking detector covers the pseudorapidity range \(|\eta| < 2.5\).
It consists of silicon pixel, silicon microstrip, and transition radiation tracking detectors.
Lead/liquid-argon (LAr) sampling calorimeters provide electromagnetic (EM) energy measurements
with high granularity.
A steel/scintillator-tile hadronic calorimeter covers the central pseudorapidity range (\(|\eta| < 1.7\)).
The endcap and forward regions are instrumented with LAr calorimeters
for EM and hadronic energy measurements up to \(|\eta| = 4.9\).
The muon spectrometer surrounds the calorimeters and is based on
three large air-core toroidal superconducting magnets with eight coils each.
The muon spectrometer includes a system of precision tracking chambers and fast detectors for triggering.
A two-level trigger system is used to select events.
The level-1 trigger is implemented in hardware and uses information from the calorimeters and the muon spectrometer
to accept events at a maximum rate of \SI{100}{\kHz}.
This is followed by a software-based high-level trigger (HLT) that
reduces the event rate to \SI{1}{\kHz} on average
depending on the data-taking conditions.
 
\section{Data and simulated event samples}
\label{sec:samples}
 
The dataset used in this analysis consists of proton--proton collision data collected with the ATLAS detector
during the second run of the LHC from 2015 to 2018
at a center-of-mass energy of \tttev
and with a minimum separation of \unit[25]{ns} between consecutive crossings of proton bunches from the two beams.
After applying data-quality requirements that ensure that all detector subsystems were operational,
the total integrated luminosity of this data sample is \ifbx{139}
with an uncertainty of \percent{1.7}~\cite{ATLAS-CONF-2019-021},
obtained using the LUCID-2 detector~\cite{LUCID2} for the primary luminosity measurements.

The SUSY signal and SM background processes are modeled with Monte Carlo (MC) simulations,
except for the multijet background,
which is estimated from data.
The modeling of the two dominant SM background processes,
namely top-quark production and production of $Z$ bosons with decays into $\tau$-leptons (\Ztautau),
was improved by normalizing their contributions to data as described in \cref{sec:background}.
Simulated samples were produced using the ATLAS simulation infrastructure~\cite{SOFT-2010-01}
with either a full simulation of the ATLAS detector in $\GEANT4$~\cite{Agostinelli:2002hh},
or a faster variant that relies on a parameterized response of the calorimeters~\cite{ATL-PHYS-PUB-2010-013}.
The latter was only used for the simulation of bottom-squark signals and to evaluate systematic uncertainties associated with generator modeling.
The effect of multiple interactions in the same and neighboring bunch crossings (pileup)
was modeled by overlaying the hard-scattering event with simulated inelastic $pp$ collisions generated with \pythia~8.186~\cite{Sjostrand:2007gs}
using the NNPDF2.3LO set of parton distribution functions (PDFs)~\cite{Ball:2012cx}
and the A3 set of tuned parameters (tune)~\cite{ATL-PHYS-PUB-2016-017}.
Simulated event samples were weighted to reproduce the distribution of the number of pileup interactions observed in the data.
For all simulated samples except those generated with \sherpa~\cite{OUR_Bothmann:2019yzt}, the \evtgen~\cite{Lange:2001uf} program
was used to simulate the decays of bottom and charm hadrons.

The production of \ttbar events was modeled using the \POWHEGBOXV{v2} generator~\cite{Frixione:2007nw,Nason:2004rx,Frixione:2007vw,Alioli:2010xd}
at next-to-leading order (NLO) in QCD with the NNPDF3.0NLO PDF set~\cite{Ball:2014uwa}
and the \hdamp\ parameter\footnote{
The \hdamp\ parameter is a resummation damping factor that controls the matching of \POWHEG matrix elements to the parton shower
and regulates the high-\pt radiation against which the \ttbar system recoils.}
set to $1.5\mtop$~\cite{ATL-PHYS-PUB-2016-020}.
Parton showering, hadronization, and the underlying event were modeled with \pythia~8.230~\cite{Sjostrand:2014zea}, using the A14 tune~\cite{ATL-PHYS-PUB-2014-021} and the NNPDF2.3LO PDF set.
The \ttbar sample was normalized to the cross-section prediction at next-to-next-to-leading order (NNLO)
in QCD, including the resummation of next-to-next-to-leading logarithmic (NNLL) soft-gluon terms calculated using
\textsc{Top++2.0}~\cite{Beneke:2011mq,Cacciari:2011hy,Baernreuther:2012ws,Czakon:2012zr,Czakon:2012pz,Czakon:2013goa,Czakon:2011xx}.
 
The production of a top quark in association with a $W$ boson was modeled using the
\POWHEGBOXV{v2}
generator~\cite{Re:2010bp,Nason:2004rx,Frixione:2007vw,Alioli:2010xd} at NLO in QCD using the five-flavor scheme.
Single-top-quark production in the $t$-channel was modeled using the
\POWHEGBOXV{v2}
generator~\cite{Frederix:2012dh,Nason:2004rx,Frixione:2007vw,Alioli:2010xd} at NLO in QCD using the four-flavor scheme.
The $s$-channel production was modeled using the \POWHEGBOXV{v2}
generator~\cite{Alioli:2009je,Nason:2004rx,Frixione:2007vw,Alioli:2010xd} at NLO in QCD in the five-flavor scheme.
For all three channels, the NNPDF3.0NLO PDF set was used for the matrix elements calculation.
The events were interfaced with \pythia~8.230 using the A14 tune and the NNPDF2.3LO PDF set.
 
Production of top-quark pairs in association with a $W$, $Z$ or Higgs boson (collectively denoted by \ttX)
was modeled using the \mgamc~v2.3.3 generator~\cite{Alwall:2014hca} at NLO in QCD with NNPDF3.0NLO PDFs.
The events were interfaced to \pythia~8.210 using the A14 tune
and the NNPDF2.3LO PDF set.
 
The production of \Vjets ($V=W,Z$) was simulated with the \sherpa~v2.2.1 generator~\cite{OUR_Bothmann:2019yzt} using NLO matrix elements for up to two jets,
and leading-order (LO) matrix elements for up to four jets calculated with the Comix~\cite{Gleisberg:2008fv}
and OpenLoops libraries~\cite{Cascioli:2011va,Denner:2016kdg}.
They were matched with the \sherpa parton showers~\cite{Schumann:2007mg} using the MEPS@NLO prescription~\cite{Hoeche:2011fd,Hoeche:2012yf,Catani:2001cc,Hoeche:2009rj}
and the tune developed by the \sherpa authors.
The NNPDF3.0NNLO PDF set~\cite{Ball:2014uwa} was used and the samples were normalized to a NNLO prediction~\cite{Anastasiou:2003ds}.

The SUSY signal samples were generated with \mgamc v2.2.3~\cite{Alwall:2014hca} using NNPDF2.3LO PDFs,
and the modeling of the parton showering, hadronization and underlying event was performed with \pythia~8.210 with the A14 tune.
The LO matrix elements include the emission of up to two additional partons.
The matching between parton showers and matrix elements was done with the CKKW-L prescription~\cite{Lonnblad:2001iq,Lonnblad:2011xx},
with a matching scale set to one quarter of the mass of the bottom squark.
Signal samples were generated with bottom-squark masses \msb ranging from \GeVx{250} to \GeVx{1000},
and masses of the second-lightest neutralino \mneuttwo between \GeVx{131} and \GeVx{380}.
Signal cross sections were calculated to approximate NNLO in QCD,
adding the resummation of soft gluon emission at NNLL accuracy~\cite{Beenakker:2016lwe,OUR_Beenakker:2014sma,Beenakker:2013mva,Beenakker:2011sf,Beenakker:2009ha,Kulesza:2009kq,Kulesza:2008jb,Beenakker:1996ch}.
The nominal cross sections and their uncertainties were derived using the PDF4LHC15\_mc PDF set,
following the recommendations of Ref.~\cite{Butterworth:2015oua},
and decrease from $24.8\pm1.6$\,pb at $\msb = \GeVx{250}$
to $14.5\pm1.5$\,fb at $\msb = \GeVx{900}$.

 
\section{Event reconstruction}
\label{sec:reconstruction}

In this section, the reconstruction of the analysis objects from the detector data is described.
The search presented in this \article is based on events
which have \bjets, hadronically decaying $\tau$-leptons and large missing transverse momentum in the final state.
In addition to these,
selections are used where $\tau$-leptons are substituted with muons
to improve the background model.
 
Inner-detector tracks with $ \pt > \MeVx{500}$ are used to reconstruct primary vertices~\cite{ATL-PHYS-PUB-2015-026}.
If several vertex candidates are found,
the one with the largest sum of the squared transverse momenta of associated tracks $\Sigma \pt^2$ is treated as the hard-scattering vertex.
 
An \antikt clustering algorithm~\cite{Cacciari:2008gp,Fastjet} with a radius parameter of $R = 0.4$ is used to reconstruct jet candidates in the calorimeter.
Jets are built from massless positive-energy topological clusters~\cite{PERF-2014-07}
of calorimeter cells containing energy above a noise threshold, measured at the electromagnetic energy scale.
The jet candidates are calibrated using jet energy scale (JES) corrections derived from data and simulation~\cite{PERF-2016-04}.
A global sequential calibration procedure is applied to improve the jet energy resolution (JER).
Jets with $\pt > \GeVx{20}$ and $|\eta| < 2.8$ are selected,
and a set of quality criteria are applied to reject jets not originating from $pp$ collisions~\cite{ATLAS-CONF-2015-029}.
To suppress jets from pileup interactions,
a jet-vertex-tagging algorithm~\cite{ATLAS-CONF-2014-018} is employed for jets with $\pt < \GeVx{120}$ and $|\eta| < 2.5$.
Jets containing $b$-hadrons are tagged as \bjets using
a boosted decision tree (BDT) algorithm
that exploits the impact parameters of tracks within the jet as well as secondary vertex information~\cite{FTAG-2018-01,ATL-PHYS-PUB-2015-022}.
The optimal working point for this analysis has an efficiency of \percent{77},
with an approximate misidentification probability of
\percent{20} for jets arising from charm quarks,
\percent{6.7} for hadronically decaying $\tau$-leptons, and
\percent{0.9} for light-flavor jets
in simulated \ttbar events.

The reconstruction of hadronically decaying $\tau$-leptons~\cite{ATL-PHYS-PUB-2015-045} is seeded by \antikt jets ($R=0.4$) built from topological clusters calibrated with a local hadronic weighting scheme~\cite{Barillari:1112035}.
The $\tau$-leptons are built from clusters and tracks found within $\Delta R=0.2$ of the seed jet axis.
The tracks are selected by a set of BDTs, and only the candidates with one or three associated tracks and a charge sum of $\pm 1$
are considered.
The $\tau$-leptons are required to have $\pt > \GeVx{20}$ and $|\eta|<2.5$, and the transition region between barrel and endcap calorimeters ($1.37 < |\eta| < 1.52$) is excluded.
The energy calibration is based on a boosted regression tree that exploits energy and shower-shape measurements from the calorimeter, information from particle-flow reconstruction~\cite{PERF-2014-06} and the number of pileup interactions.
A recurrent neural network algorithm~\cite{ATL-PHYS-PUB-2019-033} is used to distinguish between jets and $\tau$-leptons.
It uses as input a set of high-level variables combining tracking and calorimeter measurements,
as well as low-level variables from individual tracks and clusters.
The \textit{loose} identification working point is applied,
corresponding to efficiencies of \percent{85} and \percent{75} for 1-prong and 3-prong $\tau$-leptons, respectively.
To reduce background from electrons that are misidentified as $\tau$-leptons,
1-prong $\tau$-lepton candidates are discarded if a nearby electron passes the \textit{very loose} working point of the likelihood-based algorithm used to identify electrons.
This requirement is tuned to have an efficiency of \percent{95} for hadronically decaying $\tau$-leptons~\cite{ATLAS-CONF-2017-029}.
 
Muon candidates are reconstructed by combining information from the muon spectrometer and the inner tracking detectors~\cite{PERF-2015-10}.
They are required to have $\pt > \GeVx{10}$ and $|\eta| < 2.7$,
to satisfy the \textit{medium} identification criteria,
and to pass a $|z_0 \sin \theta| < \SI{0.5}{mm}$ requirement on the longitudinal impact parameter.\footnote{The transverse impact parameter is defined as the distance of closest approach in the transverse plane between a track and the beam-line.
The longitudinal impact parameter corresponds to the $z$-coordinate distance between the point along the track at which the transverse impact parameter is defined and the primary vertex.}
After discarding the candidates failing the \textit{overlap-removal} procedure described below, stricter requirements are applied:
muons must have $\pt > \GeVx{25}$, meet the \textit{loose} isolation criteria and satisfy the requirement ${|d_0|}/{\sigma(d_0)} < 3$
on the transverse impact parameter $d_0$ and its uncertainty $\sigma(d_0)$.
 
Electron candidates are reconstructed by matching energy clusters in the electromagnetic calorimeter to tracks from the inner tracking detector~\cite{PERF-2017-01}
and are required to have $\pt > \GeVx{10}$ and $|\eta| < 2.47$.
A requirement on the longitudinal impact parameter $|z_0 \sin \theta| < \SI{0.5}{mm}$ discards electrons not associated with the primary vertex.
Electrons are included in the computation of missing transverse momentum and in the overlap-removal procedure,
but are not used otherwise.
 
The missing transverse momentum vector \vecmet is defined as the negative vector sum of the transverse momenta of all reconstructed objects mentioned above,
with an additional soft term including all tracks
from the primary vertex that are not associated with a reconstructed object~\cite{ATLAS-CONF-2018-023}.
The magnitude of \vecmet is denoted by \met.
 
An overlap-removal procedure is performed after event reconstruction
to resolve ambiguities when a single physical object is reconstructed as multiple final-state objects.
If two electrons share the same track, the electron with lower transverse momentum is discarded.
Any $\tau$-leptons overlapping with an electron or a muon within $\Delta R_y \equiv \sqrt{(\Delta y)^2 + (\Delta \phi)^2} < 0.2$ are removed.
If an electron and a muon share the same inner-detector track,
the muon is removed if it is tagged as a minimum-ionizing particle in the calorimeter,
otherwise the electron is discarded.
If a jet overlaps with an electron or a muon candidate within $\Delta R_y < 0.2$, the jet is removed.
An exception is when a jet that has more than two associated tracks overlaps with a muon within $\Delta R_y < 0.2$,
in which case the jet is kept and the muon is discarded.
Finally, electron and muon candidates lying $0.2 < \Delta R_y < 0.4$ from a jet
and jets within $\Delta R_y = 0.2$ of a $\tau$-lepton candidate are discarded.
 
The same reconstruction and identification algorithms are used for both data and simulation.
Dedicated correction factors are applied to jet, $\tau$-lepton, electron and muon candidates to account for differences in efficiencies and energy calibrations between data and simulation.
 
 
\section{Event selection}
\label{sec:selection}

All selections used in this analysis require events to pass a \met trigger~\cite{METTrigger2020}
or a combined \bmet trigger~\cite{OUR_Aad:2021vwk},
except for specific selections used for the background estimate
which rely on single-muon or single-jet triggers as described in Section~\ref{sec:background}.
The \bjet and muon objects reconstructed by the trigger algorithms
are required to geometrically match the corresponding reconstructed analysis objects defined in Section~\ref{sec:reconstruction},
otherwise the event is discarded.
The HLT threshold of the \met trigger increased from $70$ to \GeVx{110} over the data-taking period.
The \bmet trigger had HLT thresholds of \GeVx{60} on \met and \GeVx{80} on the transverse momentum of the \bjet,
and the efficiency of the online \bjet identification algorithm determined for simulated \ttbar events was \percent{60} in 2016 and \percent{50} in 2017 and 2018.
This trigger increases the acceptance for low-\met signals expected from low-mass bottom squarks.
The dataset associated with the \bmet trigger has a reduced integrated luminosity of \ifbx{127}
because this trigger was not active in 2015, and stricter data-quality requirements are applied to \bjet triggers in 2016 and 2017
to ensure a valid beam-spot determination.
 
Events are rejected if no primary vertex with at least two tracks is found
or if they contain a jet failing to meet the \textit{loose} quality criteria described in Ref.~\cite{ATLAS-CONF-2015-029}.
Furthermore, events are rejected if they contain muons with a large track-curvature uncertainty
or muons which are likely to originate from cosmic rays as indicated by a large displacement from the primary vertex.
 
Events are required to have at least three jets, among which at least two must be $b$-tagged unless stated otherwise.
The leading and subleading jets are required to have $\pt > \GeVx{140}$ and $\pt > \GeVx{100}$, respectively,
and the leading \bjet is required to have $\pt > \GeVx{100}$.
The \met requirement depends on the trigger considered:
the \bmet trigger reaches maximum efficiency for $\met > \GeVx{160}$,
while the \met trigger requires $\met > \GeVx{200}$ to be fully efficient.
 
To suppress the multijet background, events are vetoed
if the angular separation in the transverse plane \dPhiTwoJetMET between one of the two leading jets and \vecmet is less than $0.5$.
All analysis selections require the presence of at least one $\tau$-lepton or one muon in the event.
This common preselection is summarized in Table~\ref{tab:plateau_cuts}.
In the following, the number of objects in an event is generically written as $N_\text{object}$,
and indices `1' and `2' refer to the leading and subleading objects, respectively,
which are ordered by decreasing transverse momentum.

\begin{table}[t]
\caption{
Summary of the common analysis preselection.
The requirements in the upper part of the table apply to all analysis regions,
those in the lower part of the table to all but the $Z(\tau\tau)$ control regions as discussed in~\cref{sec:background}.
}
\label{tab:plateau_cuts}
\centering
\begin{tabular}{llcl}
\toprule
$\Ntau+\Nmu$         &                & $\geq 1$            &                \\
\Njets               &                & $\geq 3$            &                \\
$\pt(\text{jet}_1)$  &                & $> \GeVx{140}$      &                \\
$\pt(\text{jet}_2)$  &                & $> \GeVx{100}$      &                \\
\dPhiTwoJetMET       &                & $ > 0.5$            &                \\
\midrule
\Nbjets              &                & $\geq 2$            &                \\
$\pt(\bjet_1)$       &                & $> \GeVx{100}$      &                \\
Trigger              & \bmet          & \multirow{2}{*}{OR} & \met           \\
\met                 & $> \GeVx{160}$ &                     & $> \GeVx{200}$ \\
\bottomrule
\end{tabular}
\end{table}

On top of the preselection from Table~\ref{tab:plateau_cuts},
a set of signal regions (SRs) are defined in order to target the bottom-squark signal processes illustrated in Figure~\ref{fig:sbottom_diagram}.
All SRs require at least two hadronically decaying $\tau$-leptons with opposite electric charge (referred to as the OS criterion) and no muon to be present.
 
Additional kinematic selections are applied to suppress the SM background.
These selections are described in the following and summarized in Table~\ref{tab:SR_cuts}.
They are optimized by maximizing the signal significance~\cite{Cowan:2010js}
in the previously nonexcluded parameter space of the targeted signal model.
 
To ensure compatibility with a Higgs boson decay,
the visible invariant mass of the two leading $\tau$-leptons must satisfy $\GeVx{55} < \minvtautau < \GeVx{120}$.
The lower bound suppresses the \Ztautau background, while the upper bound reduces `nonresonant' background contributions
where the $\tau$-leptons do not originate from the same resonance.
Events are required to have $\HT > \GeVx{1100}$, where
$\HT \equiv\sum \pt^{\mathrm{\tau}}+ \sum \pt^{\mathrm{\mu}}+ \sum \pt^{\mathrm{jet}}$
is the scalar sum of the transverse momenta of all $\tau$-leptons, muons and jets in the event.
This variable exploits the fact that signals with large bottom-squark masses are expected to produce highly boosted particles in the final state.

The stransverse mass variable~\cite{Lester:1999tx,Lester:2014yga}, denoted \mttwo,
is used to discriminate between the signal process and the top-quark production background.
It is designed to have an endpoint
for background processes such as top-quark production
where the two $\tau$-leptons originate from separate decay branches.
For the signal process, the two $\tau$-leptons originate from a resonant Higgs boson decay,
and the \mttwo spectrum has a pronounced tail towards larger values.
The \mttwo variable is computed as 
\begin{equation}\nonumber
\mttwo = \min_{\vec{p}_\text{T}^{a} + \vec{p}_\text{T}^{b} = \vecmet} \left(\max\left[\mt(\vec{p}_{\text{T}}^{\,\tau_1},\vec{p}_\text{T}^{\,a}) , \mt(\vec{p}_{\text{T}}^{\,\tau_2},\vec{p}_\text{T}^{\,b})   \right]\right),
\end{equation}
where $\vec{p}_\text{T}^{\,\tau_1,\tau_2}$ correspond to the transverse momenta of the two leading $\tau$-leptons,
and $(a,b)$ refers to two invisible particles assumed to be produced with transverse momentum $\vec{p}_\text{T}^{\,a,b}$.
The masses of the invisible particles are free parameters and set to $m_a = m_b \equiv m_\text{inv}$.
The transverse mass \mt is defined as
$\mt^2(\vec{p}_{\text{T}}^{\,\tau_1}, \vec{p}_{\text{T}}^{\,a})
= m_{a}^2 + 2\left(p_\text{T}^{\,\tau_1}\sqrt{m_{a}^2+\left(p_\text{T}^{a}\right)^2}-\vec{p}_\text{T}^{\,\tau_1} \cdot \vec{p}_\text{T}^{\,a}\right)$,
where the $\tau$-lepton mass is set to \GeVx{0}.
The \mttwo distribution peaks at \GeVx{0} for both the bottom-squark signal and the dominant \ttbar background
when setting $m_\text{inv}$ to \GeVx{0}, providing poor discrimination.
The discrimination improves as $m_\text{inv}$ is increased,
and a value of \GeVx{120} is found
to result in an \mttwo distribution that best separates the signal from the background.
All SRs require $\mttwo > \GeVx{140}$.
Some of the CRs also make use of the transverse mass of a $\tau$-lepton,
which is computed as $\left(\mT^{\tau}\right)^2 = 2 ( \pt^\tau \, \met - \vec{p}_\text{T}^{\,\tau} \cdot \vecmet)$.
 
The last discriminant is \mintheta,
defined as the smallest three-dimensional angle
of the four combinations between either of the two leading $\tau$-leptons and either of the two leading \bjets.
For the \ttbar~background, the smallest angle is expected from configurations where the \bjet and the $\tau$-lepton originate from the same top-quark decay,
resulting in relatively low values of \mintheta.
For $Z(\tau\tau)+b\bar b$ events with a highly boosted $Z$ boson, the pair of $\tau$-leptons recoils against the \bjets, and large values of \mintheta are expected.
For signal events where $\sbottom \to b \ninotwo \to b h(\tau\tau) \ninoone$,
the angle between the \bjet and the $\tau$-lepton pair increases with the \sbottom mass, and so does \mintheta.
A multi-bin SR with three \mintheta bins ($< 0.5, [0.5,1.0], >1.0$) is defined in order to take advantage of these features.
A single-bin SR requiring $\mintheta > 0.6$ is used to provide cross-section limits on generic processes beyond the Standard Model (BSM).
The probability for a signal event to enter the single-bin SR ranges between
$6.4\cdot10^{-6}$ at $\msb = \GeVx{250}$ and $\mneuttwo = \GeVx{150}$ and
$1.4\cdot10^{-3}$ at $\msb = \GeVx{900}$ and $\mneuttwo = \GeVx{150}$,
taking into account the Higgs boson and $\tau$-lepton branching ratios, the SR acceptance, and particle reconstruction and identification efficiencies.
The requirement responsible for the largest decrease in signal acceptance is the presence of two hadronically decaying $\tau$-leptons in the final state.
 
Examples of signal and background kinematic distributions are shown in Figure~\ref{fig:preselection}.
The three plots show the \HT, \minvtautau and \mttwo variables after the preselection.
The estimated \SM background is scaled by the normalization factors from the background fit described in Section~\ref{sec:background},
and the distributions for several signal models are overlaid.

\begin{table}[t]
\caption{
Definition of the single-bin and multi-bin signal regions.
The requirements are applied in addition to the preselection from Table~\ref{tab:plateau_cuts}.
The single-bin and multi-bin SRs only differ by the \mintheta requirement.
}
\centering
\begin{tabular}{lcc}
\toprule
& \multicolumn{2}{c}{Common SR requirements}             \\
\midrule
\Nmu                  & \multicolumn{2}{c}{$0$}                                \\
\Ntau                 & \multicolumn{2}{c}{$\geq 2$}                           \\
OS$(\tau_1, \tau_2)$  & \multicolumn{2}{c}{yes}                                \\
\minvtautau           & \multicolumn{2}{c}{\,$[55, 120]$~\gev}                   \\
\mttwo                & \multicolumn{2}{c}{~~$> \GeVx{140}$}                     \\
\HT                   & \multicolumn{2}{c}{$> \GeVx{1100}$}                    \\
\midrule
& Single-bin SR & Multi-bin SR                           \\
\midrule
\mintheta             & $> 0.6$       & 3 bins: $< 0.5$, $[0.5,1.0]$, $>1.0$   \\
\bottomrule
\end{tabular}
\label{tab:SR_cuts}
\end{table}
 
\begin{figure}[t]
\centering
\subfloat[]{\includegraphics[width= 0.325\textwidth]{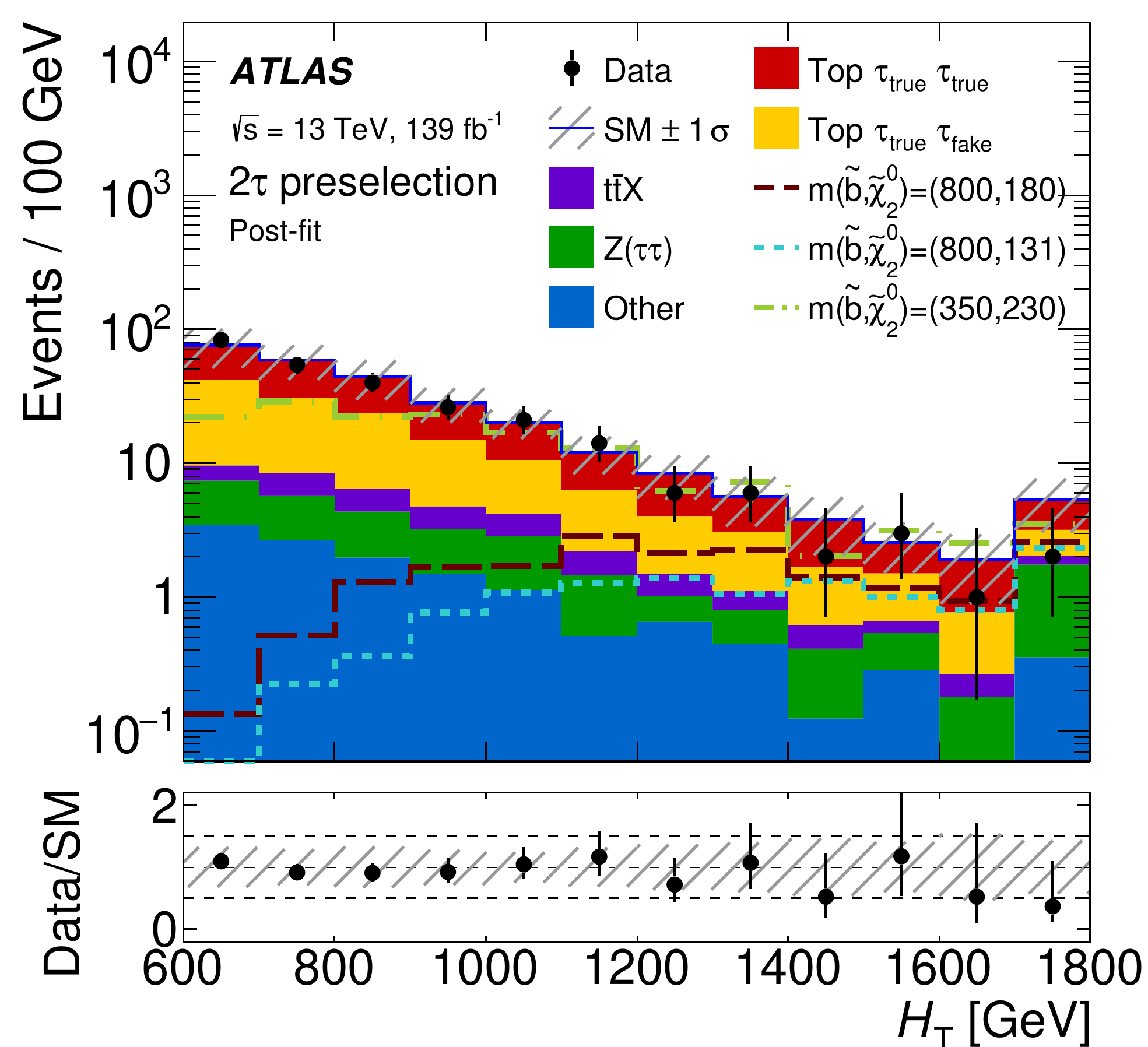} \label{fig:SR_ht_afterFit}}
\hfill
\subfloat[]{\includegraphics[width= 0.325\textwidth]{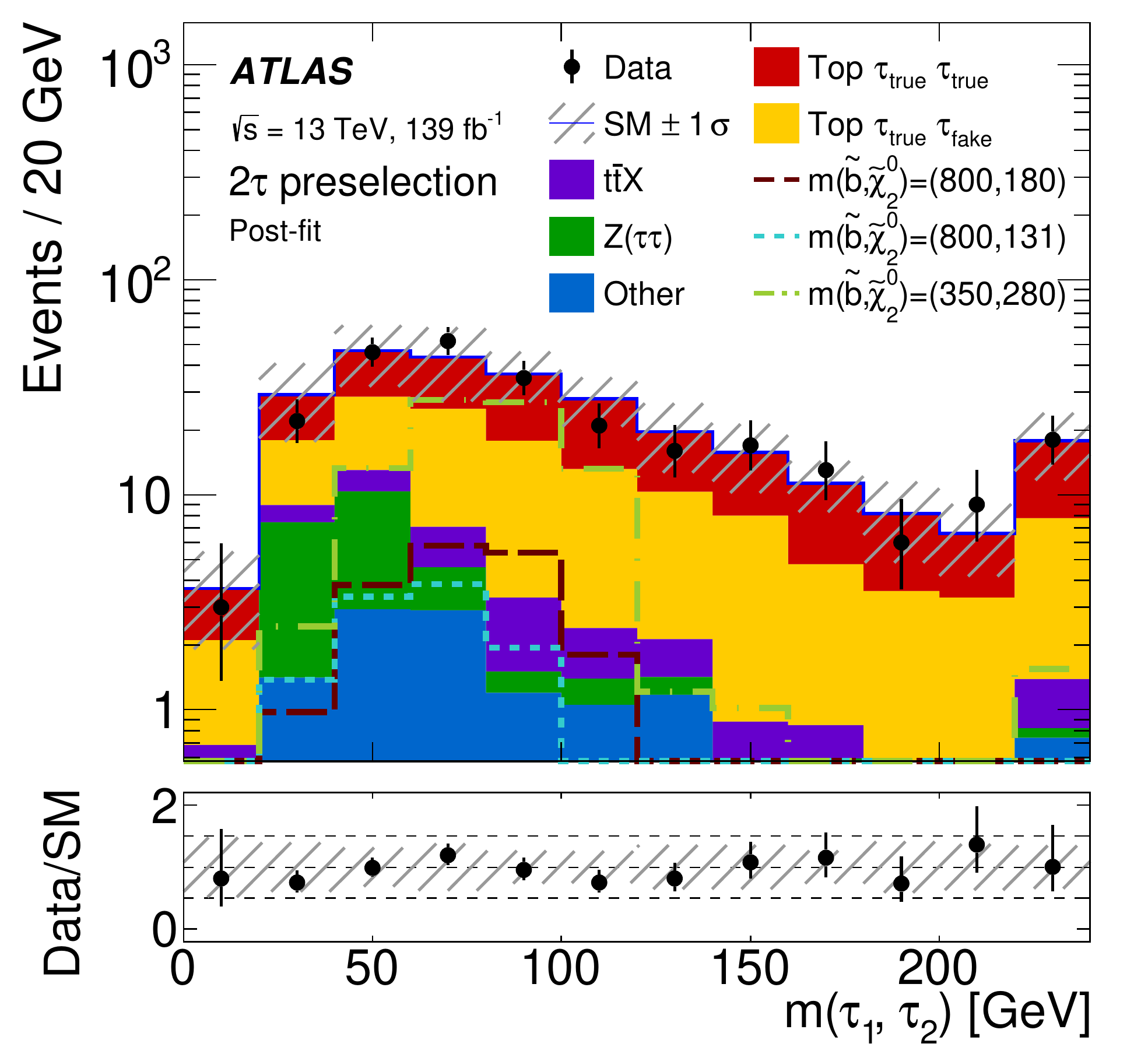} \label{fig:SR_ditau_mass_afterFit}}
\hfill
\subfloat[]{\includegraphics[width= 0.325\textwidth]{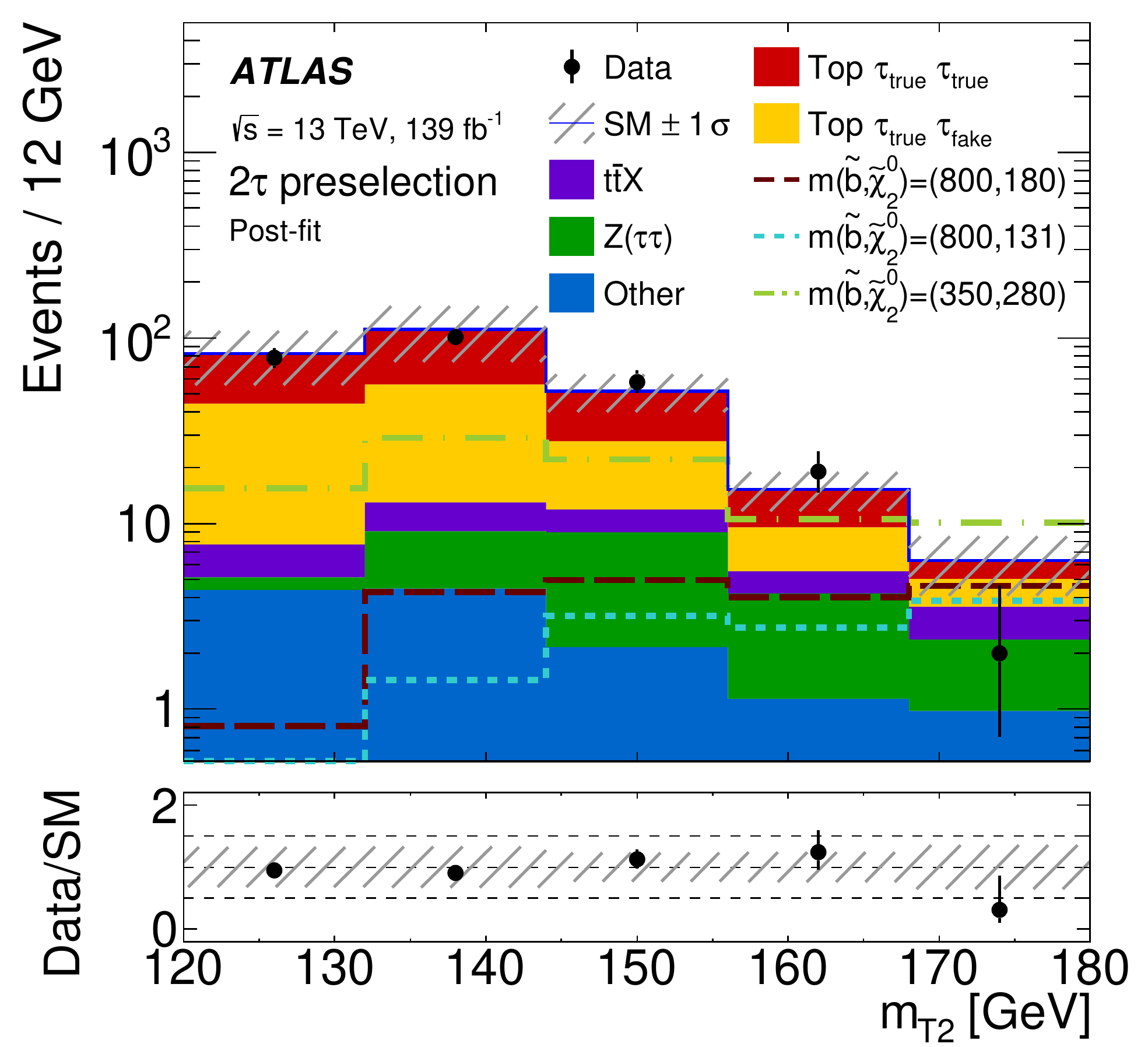} \label{fig:SR_tau_mt2_afterFit}}
 
\caption{
Kinematic distributions of data and SM background for events that pass the preselection and have at least two hadronically decaying $\tau$-leptons.
Predictions from three signal models are also shown, where the masses \msb and \mneuttwo are given in \gev\ in the legend.
Distributions are displayed for the \protect\subref{fig:SR_ht_afterFit} \HT, \protect\subref{fig:SR_ditau_mass_afterFit} \minvtautau and \protect\subref{fig:SR_tau_mt2_afterFit} \mttwo variables.
\BandIsStatSyst
The ``Other'' contribution includes all the backgrounds not explicitly listed in the legend (\Vjets except \Ztautaujets, diboson/triboson, multijet).
\PostfitRef
\RightBinOverflow
\BottomPanelDataSM
}
\label{fig:preselection}
\end{figure}

\FloatBarrier
 
\section{Background estimation}
\label{sec:background}

The largest backgrounds in the SRs are from  \ttbar and single-top-quark processes, referred to as top-quark background, and \Ztautau produced in association with \bjets.
Subdominant contributions arise from \ttX~processes,
while other backgrounds such as multijet or diboson and triboson production are found to be negligible.
The normalization of the two dominant backgrounds is fitted to the data in dedicated control regions (CRs)
kinematically close to the SRs but where little signal is expected.
The normalization factors are derived with a likelihood fit based on the \histfitter framework~\cite{Baak:2014wma}.
The fit uses as input the observed data yields, the expected yields predicted from simulation, as well as the statistical and systematic uncertainties described in Section~\ref{sec:systematics}.
Two main fit setups are employed in the analysis.
The background-only fit refers to the configuration that only includes the CRs, and where no signal is considered.
The signal-plus-background fit includes both the CRs and the SRs, and it takes into account a possible signal contribution in the fitted regions.
It is used to establish exclusion limits as discussed in Section~\ref{sec:result}.
In both cases, the fit is performed simultaneously over all the relevant regions.
Subdominant background contributions are normalized according to their cross sections and the integrated luminosity of the data.
The multijet background is determined from data.
Validation regions (VRs) are defined in phase-space regions as close as possible to that of the SRs.
The VRs are not included in the fit.
They are used to validate the background-model extrapolation from the CRs to the SRs by comparing the observed data with the fitted background predictions.
As such, they are designed to have little signal contribution.
The methods used to estimate the various backgrounds are described in the following, together with the associated CRs and VRs.

Multijet production is an important background at hadron colliders, but it is efficiently suppressed in this analysis
by the requirement of two hadronically decaying $\tau$-leptons, two \bjets, large \met, and $\dPhiTwoJetMET > 0.5$.
A data-driven jet-smearing method~\cite{SUSY-2011-20} is employed to estimate this background.
Events recorded by single-jet triggers are processed through an energy-smearing procedure that emulates \met originating from resolution effects.
The normalization of the smeared pseudo-data template is derived in events where one of the two leading jets is aligned with \vecmet in the transverse plane.
Except for that multijet-enriched selection, the multijet background is found to be negligible in all analysis selections.
Therefore, its normalization is kept constant in the fits, for simplicity.

The design of the control regions for the top-quark and \Ztautaubbbar backgrounds is driven by two main considerations.
Firstly, the hadronically decaying $\tau$-leptons selected in the analysis are either prompt $\tau$-leptons from electroweak boson decays,
or jets misidentified as $\tau$-leptons.
They are referred to as \textit{true} $\tau$-leptons (\ttrue) and \textit{fake} $\tau$-leptons (\tfake), respectively,
and their contributions must be handled separately in the background model.
No such distinction is made for \bjets, as the fraction of misidentified \bjets does not exceed \percent{10} in the analysis phase space.
The top-quark background in the SRs is composed of $\ttrue\ttrue$ and $\ttrue\tfake$ contributions of comparable magnitude,
where one $\tau$-lepton comes from a $W$-boson decay, and the second $\tau$-lepton either comes from the other $W$-boson decay or from a jet misidentified as a $\tau$-lepton.
The $\tfake\tfake$ contribution is negligible due to the large jet rejection provided by the $\tau$-lepton identification algorithm.
In the case of \Ztautaubbbar events, only the $\ttrue\ttrue$ contribution is found to be relevant.
Secondly, the background normalization factors cannot be accurately determined using events containing two hadronically decaying $\tau$-leptons (\tauhad) and two \bjets,
as the low event yields remaining after the preselection
do not allow control regions with sufficient statistical power, high purity, and low signal contamination to be defined.
 
Because of these limitations
the CRs are based on final states where either one or two $\tau$-leptons are replaced with muons.
The \CRTrueTau and \CRFakeTau selections are defined to respectively target top-quark events
with one muon plus either one \ttrue or one \tfake in the final state,
where the muon replaces a \ttrue from one of the $W$-boson decays.
The \CRZmmbb region is defined to select \Zmumubbbar events.
By trading $W(\tau\nu)$ for $W(\mu\nu)$ and \Ztautau for \Zmumu,
the CRs target the desired background processes
but benefit
from larger yields due to the branching ratio ${\cal B}(\tau\to\tauhad\nu_\tau)$ of \percent{65} that does not apply to muons,
and the reconstruction and identification efficiencies that are higher for muons.
In the top-quark CRs, event yields are further increased by a combinatorial factor of two.

The normalization factors derived for background events with muons are not directly applicable to background events in the SRs
that contain two hadronically decaying $\tau$-leptons.
The replacement of $\tau$-leptons with muons has an impact on the reconstructed event kinematics and the selection efficiency of background processes,
which needs to be accounted for.
This is done by introducing additional CRs and normalization factors, two for the top-quark background and two for the \Ztautaubbbar background,
that allow an extrapolation from muon to $\tau$-lepton selections.
As mentioned in Section~\ref{sec:reconstruction},
corrections are already applied to muons and $\tau$-leptons in the simulation to match the efficiencies and energy calibration measured in data.
The background normalization factors from the additional CRs
thus mostly account for acceptance effects.

\begin{table}[t]
\caption{
Definition of the control regions used for the top-quark background.
The requirements are applied in addition to the preselection.
\DashNoCut
}
\label{tab:TopCRs_def}
\centering
\begin{tabular}{lcccc}
\toprule
& \CROneMu & \CROneTau     & \CRTrueTau    & \CRFakeTau     \\
\midrule
\Nmu           & 1        & 0             & 1             & 1              \\
\Ntau          & 0        & 1             & 1             & 1              \\
OS($\mu,\tau$) & --       & --            & yes           & yes            \\
\HT            & \multicolumn{4}{c}{$[600, 1000]$~\gev}                    \\
$\mT^{\tau}$   & --       & $< \GeVx{80}$ & $< \GeVx{80}$ & $> \GeVx{100}$ \\
\bottomrule
\end{tabular}
\end{table}

The definitions of the four control regions
used to normalize the top-quark background are summarized in Table~\ref{tab:TopCRs_def}.
The \CRTrueTau and \CRFakeTau regions select events that
contain exactly one muon and one $\tau$-lepton of opposite electric charge.
Like all control regions defined in this analysis,
they use the \HT range from 600 to \GeVx{1000}.
For \CRTrueTau, the $\tau$-lepton transverse mass $\mT^{\tau}$ must be lower than \GeVx{80},
which results in a high purity of true $\tau$-leptons.
For \CRFakeTau, $\mT^{\tau}$ has to be larger than \GeVx{100},
which gives a roughly equal mix of true and fake $\tau$-leptons.
The \CROneTau selection is identical to that of \CRTrueTau except that events must not contain a muon.
This region has a high purity in top-quark background events decaying semileptonically with a true $\tau$-lepton in the final state.
The \CROneMu selection is defined in a similar way, with one muon and no $\tau$-lepton,
selecting high-purity semileptonic top-quark processes with a muon in the final state.

\begin{figure}[t]
\centering
\subfloat[]{\includegraphics[width= 0.45\textwidth]{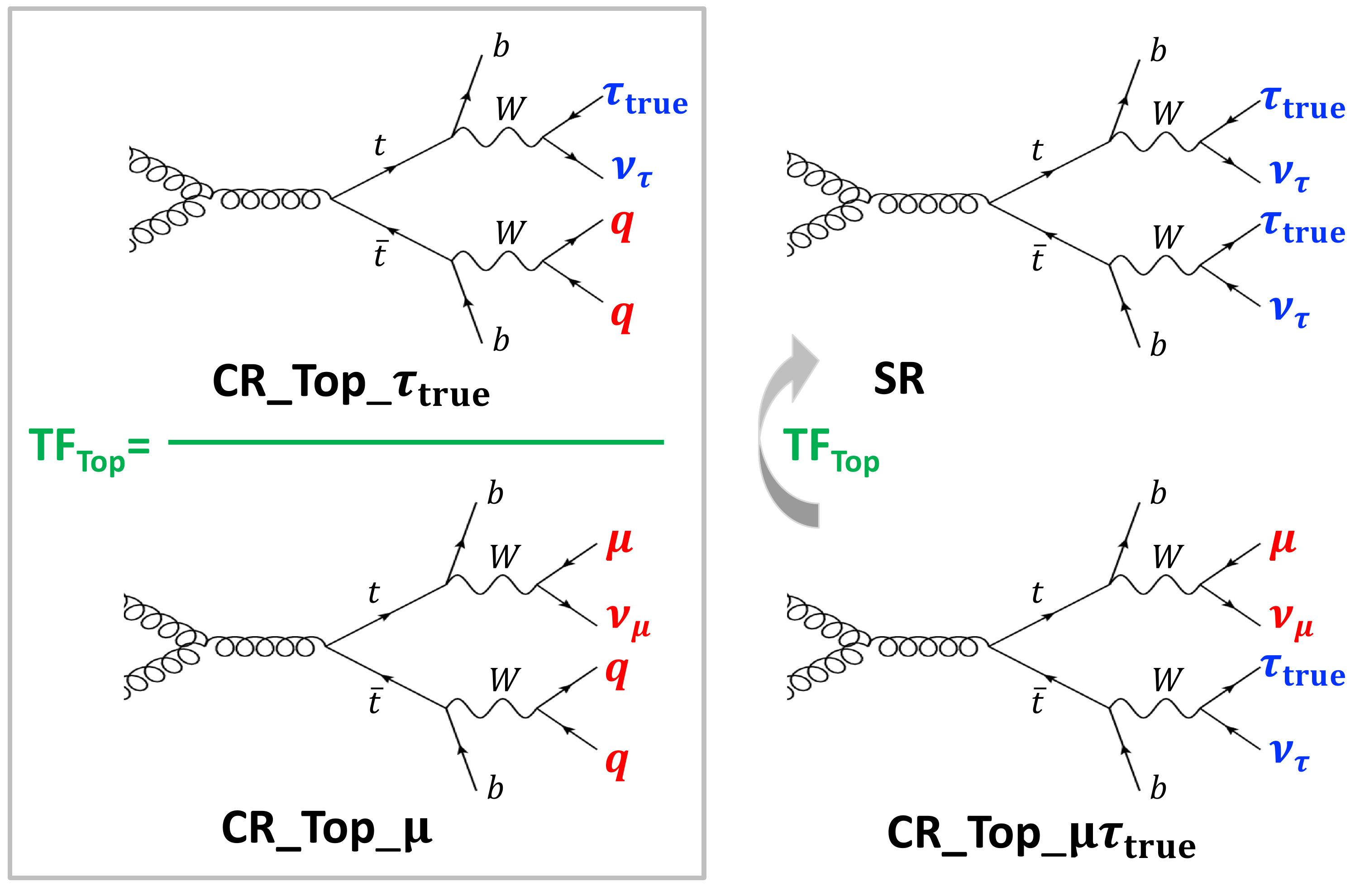}\label{fig:SchemeTopCRs}}
\hspace{1cm}
\subfloat[]{\includegraphics[width= 0.45\textwidth]{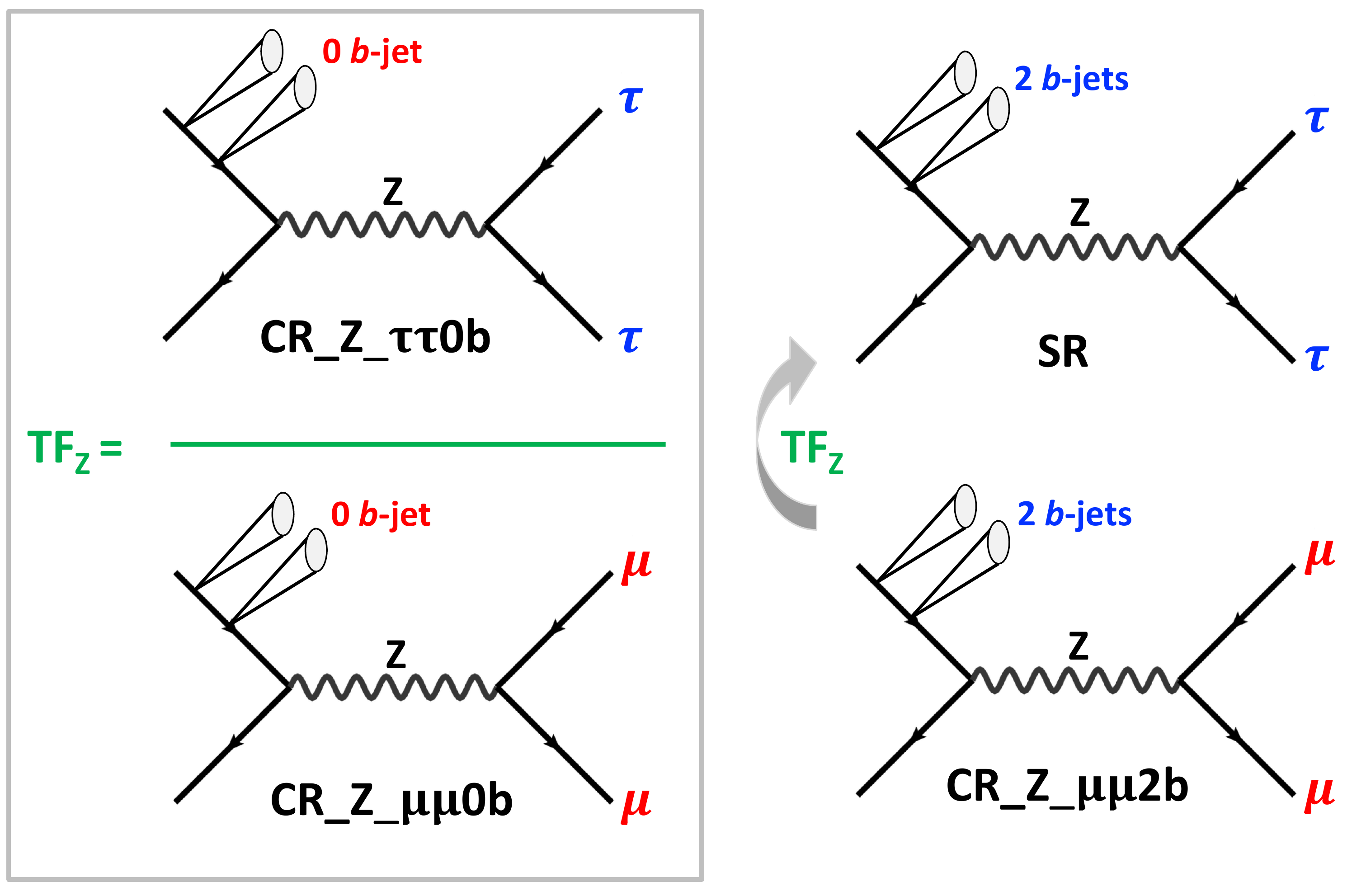}\label{fig:SchemeZCRs}}
\caption{
Schematic representation of the control region setup that is used to constrain the normalization of the
\protect\subref{fig:SchemeTopCRs}~top-quark and
\protect\subref{fig:SchemeZCRs} \Ztautaujets backgrounds in the signal regions.
The arrows represent the transfer factors associated with the replacement of muons with true $\tau$-leptons, which correct for acceptance.
For the top-quark background, the sketch illustrates the normalization strategy for the \ttrue\ttrue contribution.
A similar strategy is employed for the \ttrue\tfake contribution, where the \tfake can originate from a jet from a hadronically decaying $W$ boson, a \bjet, or a jet from initial-state radiation.
}
\end{figure}

The way the four CRs from Table~\ref{tab:TopCRs_def} are used to derive normalization factors for the top-quark background processes
is illustrated in Figure~\ref{fig:SchemeTopCRs}.
The expected yields for top-quark production with true and fake $\tau$-leptons from Monte Carlo simulation
are respectively multiplied by normalization factors \NFtoptrue and \NFtopfake,
that float freely in the fit
and are constrained by data mainly through \CRTrueTau and \CRFakeTau.
To account for the different lepton flavors in the signal region (with two $\tau$-leptons) and the control region (one $\tau$-lepton and one muon),
the top-quark production yields are further multiplied by additional freely floating normalization factors \NFonetau and \NFonemu,
which are constrained mainly through the regions \CROneTau and \CROneMu.
A transfer factor $\TFtop \equiv \NFonetau / \NFonemu$ is used to correct for the difference between requiring a muon and a true $\tau$-lepton.
This means that a simulated top-quark event with one true and one fake $\tau$-lepton in one of the signal regions
receives a normalization factor $\NFtopfake \times \TFtop$,
and a simulated top-quark event with two true $\tau$-leptons
a normalization factor $\NFtoptrue \times \TFtop$.

\begin{figure}[!ht]
\centering
\subfloat[\met in \CRTrueTau]{\includegraphics[width=0.47\textwidth]{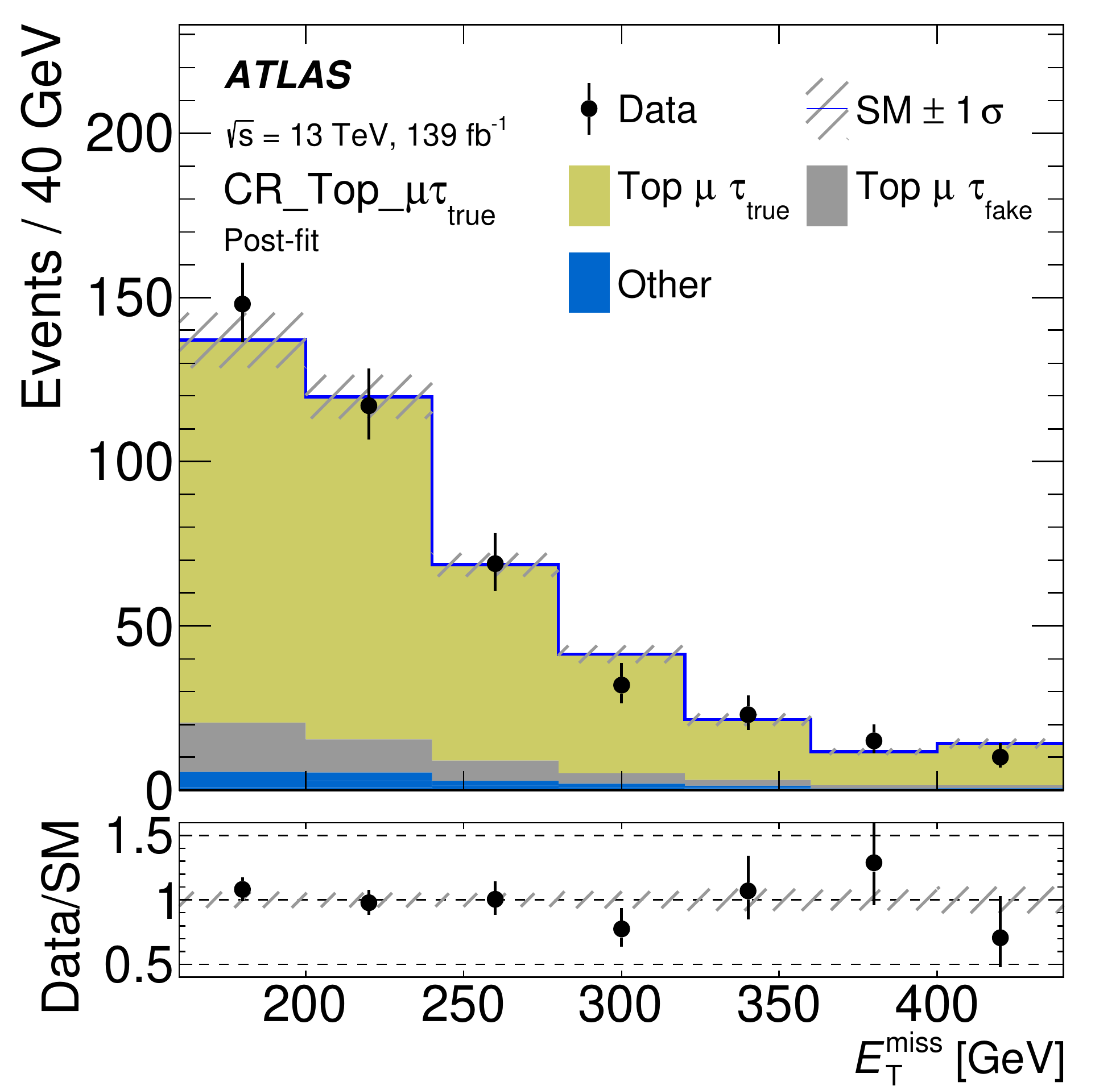} \label{fig:TrueTauCR_tau_pt}}
\hfil
\subfloat[$\pt(\tau)$ in \CRFakeTau]{\includegraphics[width=0.47\textwidth]{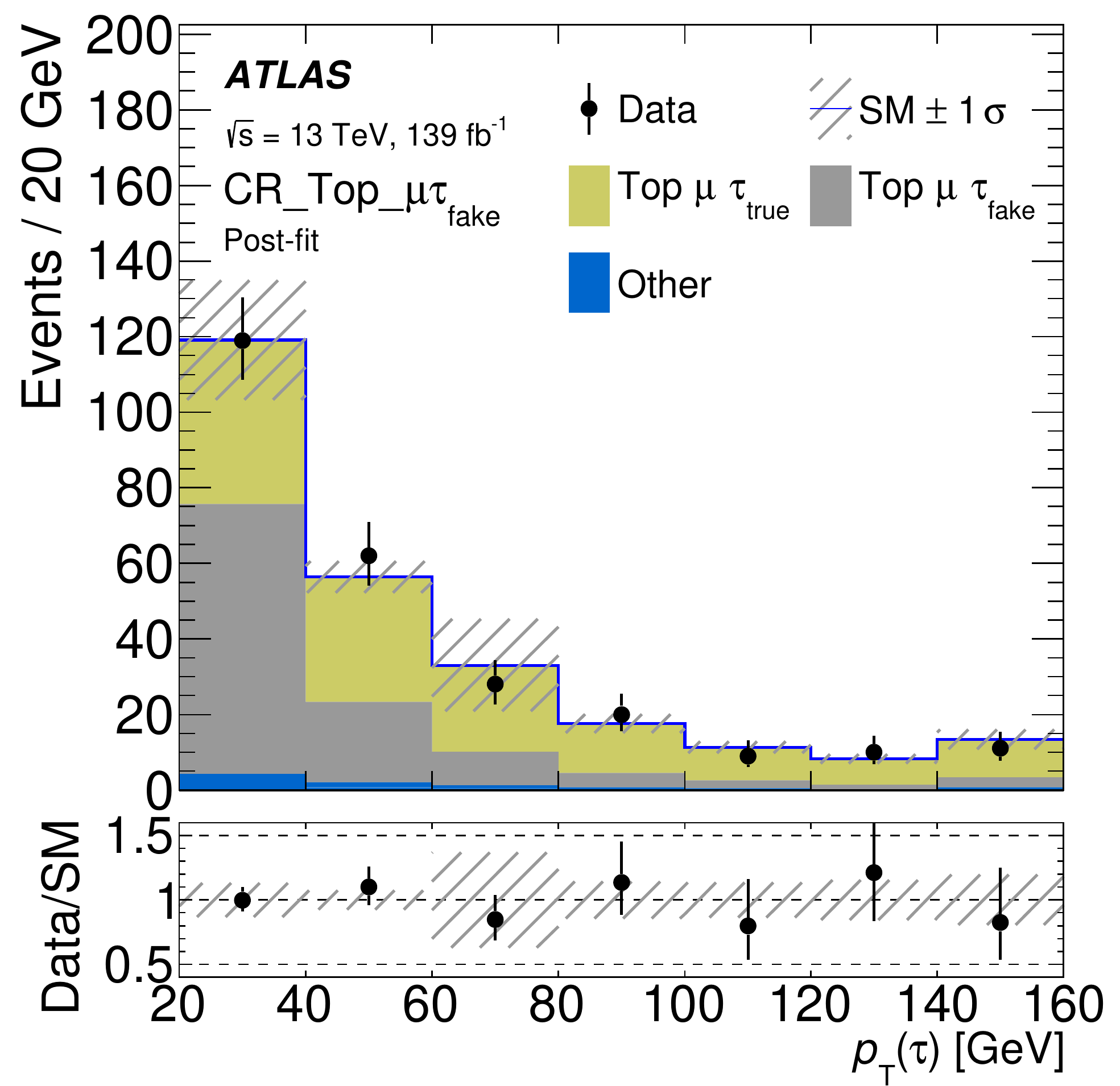} \label{fig:FakeTauCR_tau_pt}}\\
\subfloat[$\pt(\text{jet}_1)$ in \CROneMu]{\includegraphics[width=0.47\textwidth]{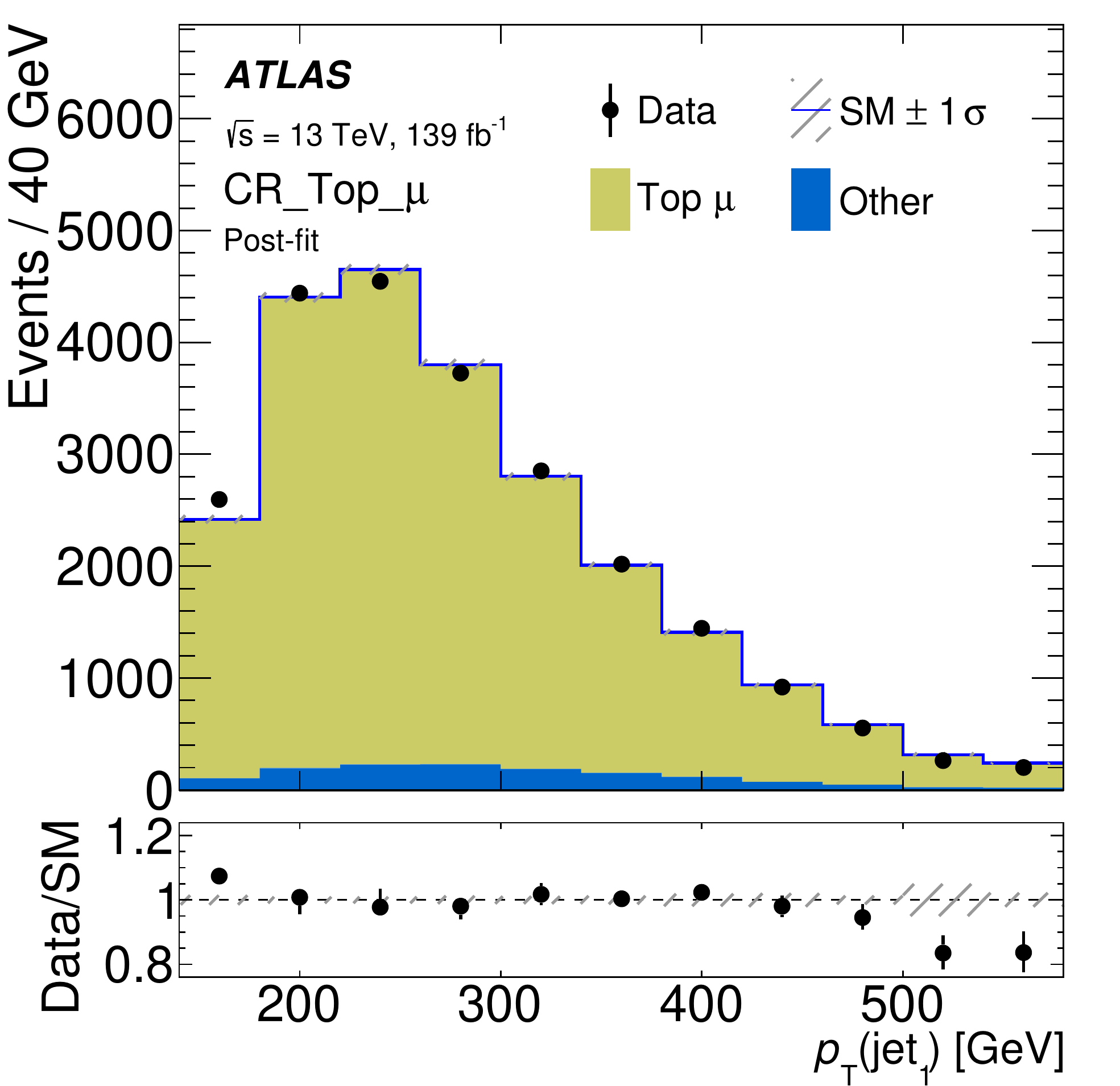} \label{fig:OneMuCR_ht}}
\hfil
\subfloat[$\mT^{\tau}$ in \CROneTau]{\includegraphics[width=0.47\textwidth]{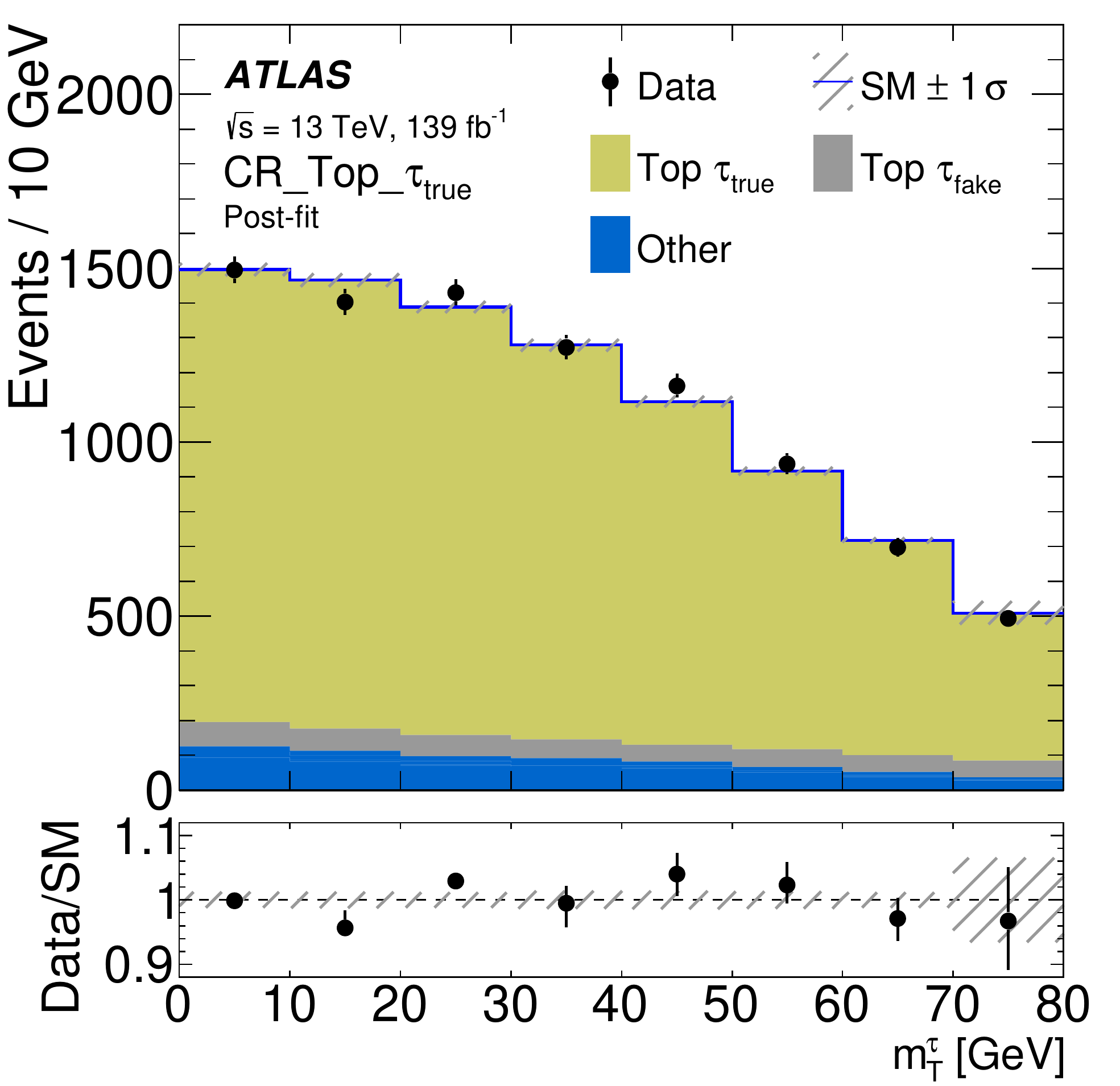}  \label{fig:OneTauCR_ht}}
\caption{
Kinematic distributions from the four control regions associated with the top-quark background, showing
\protect\subref{fig:TrueTauCR_tau_pt} $\met$ in \CRTrueTau,
\protect\subref{fig:FakeTauCR_tau_pt} $\pt(\tau)$ in \CRFakeTau,
\protect\subref{fig:OneMuCR_ht}       $\pt(\text{jet}_1)$ in \CROneMu, and
\protect\subref{fig:OneTauCR_ht}      $\mT^{\tau}$ in \CROneTau.
\BandIsStatSyst
\Postfit
The ``Other'' contribution includes all the backgrounds not explicitly listed in the legend (\Vjets, \ttX, diboson/triboson, multijet).
\RightBinOverflow
\BottomPanelDataSM
}
\label{fig:TopCRs}
\end{figure}

Figure~\ref{fig:TopCRs} shows several examples of distributions from the four control regions associated with the top-quark background.
In these plots, the predicted background contributions from simulation are scaled with the normalization factors obtained from the background-only fit.
All of the CRs show good agreement between the \SM prediction and the data.
They also have high purity in the respective top-quark background processes
except for \CRFakeTau,
where the purity is only \percent{43}
because it is difficult to isolate the contribution of the top-quark background with fake $\tau$-leptons.
 
\begin{table}[t]
\caption{
Definition of the control regions used for the \Zjets background.
The requirements are applied in addition to the set of preselection criteria reported in the upper part of Table~\ref{tab:plateau_cuts}.
\DashNoCut
}
\label{tab:ztautau_defs}
\centering
\begin{tabular}{lcccc}
\toprule
& \CRZmmbb   & \CRZmm                  & \CRZtt \\
\midrule
Trigger                       & \multicolumn{2}{c}{single-muon}      & \met         \\
\Nmu                          & \multicolumn{2}{c}{2}                & 0            \\
\Ntau                         & \multicolumn{2}{c}{0}                & 2            \\
\Nbjets                       & 2          & 0                       & 0            \\
$\pt(\mu_1)$                  & \multicolumn{2}{c}{~$> \GeVx{30}$}    & --           \\
\met                          & \multicolumn{2}{c}{$< \GeVx{100}$~~}   & $> \GeVx{200}$ \\
$\pt(\mu_1,\mu_2)$            & \multicolumn{2}{c}{$> \GeVx{200}$~~}   & --           \\
\minvmumu                     & \multicolumn{2}{c}{$[81, 101]$~\gev~~~} & --           \\
$\mT^{\tau_1} + \mT^{\tau_2}$ & \multicolumn{2}{c}{--}               & $< \GeVx{100}$ \\
\HT                           & \multicolumn{3}{c}{$[600, 1000]$~\gev}              \\
\bottomrule
\end{tabular}
\end{table}

The three control regions that target the \Ztautau background are summarized in Table~\ref{tab:ztautau_defs}.
The \CRZmmbb selection is defined using events with two muons of opposite electric charge, taken as proxies for two true $\tau$-leptons,
and two \bjets.
Since \Zmumujets processes do not have large \met in the final state, the events are selected using a single-muon trigger,
which has its efficiency plateau at $\pt(\mu) > \GeVx{30}$.
The invariant mass of the dimuon system is required to be within \GeVx{10} of the $Z$-boson mass,
and \met to be lower than \GeVx{100} to increase the purity of the selection.
To move the CR closer to the relevant phase space, \HT must be in the range $[600, 1000]~\gev$,
and the transverse momentum of the muon pair $\pt(\mu_1,\mu_2)$ must be larger than $\GeVx{200}$,
which is a typical value found in simulation for the \pt of the $Z$~boson in \Ztautau events after the preselection.
The \Zmumu background is multiplied by the freely floating normalization factor $\NFzmumubb$,
which is constrained through \CRZmmbb.
 
The two additional control regions \CRZmm and \CRZtt
are used to correct for the difference in acceptance and efficiency
when replacing the $\tau$-leptons with muons to estimate the \Zjets background.
The interplay of these CRs is illustrated in Figure~\ref{fig:SchemeZCRs}.
The \CRZmm selection is the same as for \CRZmmbb but with a \bjet veto,
whereas \CRZtt requires the presence of two $\tau$-leptons with opposite electric charge and no \bjet.
The \CRZtt events are selected with an \met trigger and $\met > \GeVx{200}$ as is done for the SRs,
and muons are vetoed in this region.
Additionally, the sum of $\tau$-lepton transverse masses $\mT^{\tau_1} + \mT^{\tau_2}$ has to be lower than \GeVx{100} to increase the purity in \Ztautau events.
In all of these three CRs, \HT is again required to be within $[600,1000]~\gev$.
 
From these two auxiliary control regions, the freely floating normalization factor \NFzmumu and transfer factor $\TFZ \equiv \NFztautau / \NFzmumu$ are derived in the background fit.
The background normalization in \CRZmm is absorbed into \NFzmumu.
The transfer factor \TFZ transfers the normalization from \CRZmm to \CRZtt,
and from \CRZmmbb to the SRs,
\ie \Ztautaubbbar events in the SRs are scaled by $\NFzmumubb \cdot \TFZ$.

\begin{table}[t]
\caption{
Values of normalization and transfer factors with their statistical and systematic uncertainties as obtained from the background-only fit,
in the top part of the table for top-quark background processes, and
in the bottom part for \Zjets.
The control regions that primarily affect the normalization factors are listed,
together with the purity of the CR in the relevant background process.
As \TFtop and \TFZ are ratios of two normalization factors,
one of which (the denominator) is listed in the row directly above,
the table lists the respective second control region (the numerator of the ratio) and its purity in top-quark or \Ztautaubbbar events.
}
\label{tab:normfactors}
\centering
\begin{tabular}{cclc}
\toprule
Normalization / transfer factor            & Fitted value                      & Control region & Purity \\
\midrule
\multirow{2}{*}{\NFtoptrue}                & \multirow{2}{*}{0.88 $\pm$  0.16} & \CRTrueTau       & $86\%$          \\
&                                   & \CRFakeTau       & $53\%$          \\
\multirow{2}{*}{\NFtopfake}                & \multirow{2}{*}{0.79 $\pm$  0.30} & \CRFakeTau       & $43\%$          \\
&                                   & \CRTrueTau       & ~~$9\%$           \\[+1mm]
\NFonemu                                   & 0.91 $\pm$  0.10                  & \CROneMu         & $94\%$          \\
$\TFtop \equiv \NFonetau / \NFonemu$       & 0.98 $\pm$  0.04                  & \CROneTau        & $88\%$          \\
\midrule
\NFzmumubb                                 & 1.28 $\pm$  0.12                  & \CRZmmbb         & $89\%$          \\
\NFzmumu                                   & 1.00 $\pm$  0.05                  & \CRZmm           & $96\%$          \\
$\TFZ \equiv \NFztautau / \NFzmumu$        & 0.99 $\pm$  0.17                  & \CRZtt           & $79\%$          \\
\bottomrule
\end{tabular}
\end{table}
 
All normalization and transfer factors are obtained from a simultaneous fit of the seven CRs for the top-quark and \Ztautau backgrounds.
Table~\ref{tab:normfactors} lists the values of the normalization factors and transfer factors and their uncertainties,
the names of the control regions that
determine the normalization factors
and the respective purities of the control regions
in top-quark or \Zjets events.
The transfer factors \TFtop and \TFZ are computed from ratios of two normalization factors
as explained above.
For these, one row in the table (\NFonemu and \NFzmumu)
gives the values forming the respective denominators of the ratios,
showing how well the data and simulated events agree in these regions.
The row below gives the transfer factor (\TFtop and \TFZ, respectively).
In these rows, the table lists the second control region (the numerator of the ratio) and its purity.

Three validation regions are defined to check the extrapolation from \CRTrueTau, \CRFakeTau, and \CRZmmbb in the \HT variable.
This is done by changing the requirement on \HT that is applied in the CRs
from $\GeVx{600} < \HT < \GeVx{1000}$ to $\GeVx{1000} < \HT < \GeVx{1500}$ in the VRs,
while keeping all other requirements the same as for the respective CRs.
Shifting the \HT range moves the validation regions closer to the signal regions,
which require $\HT > \GeVx{1100}$.
The VRs and the SRs are mutually exclusive due to the muon veto
that is part of the signal-region selections.
The names of the three VRs match those of the corresponding CRs.
A fourth validation region, \VRTop, is defined to validate the extrapolation from muons to $\tau$-leptons
in events with two \bjets and two hadronically decaying $\tau$-leptons
which pass the \met trigger or the \bmet trigger and the corresponding trigger-plateau requirements.
To avoid overlap of this VR with the SRs,
\HT is required to be within $[600,1000]~\gev$.
In addition, the visible di-$\tau$ mass \minvtautau is required to be either lower than $\GeVx{40}$ or larger than $\GeVx{90}$
to reduce the contribution from a possible bottom-squark signal.

\begin{figure}[t]
\centering
\includegraphics[width=0.9\textwidth]{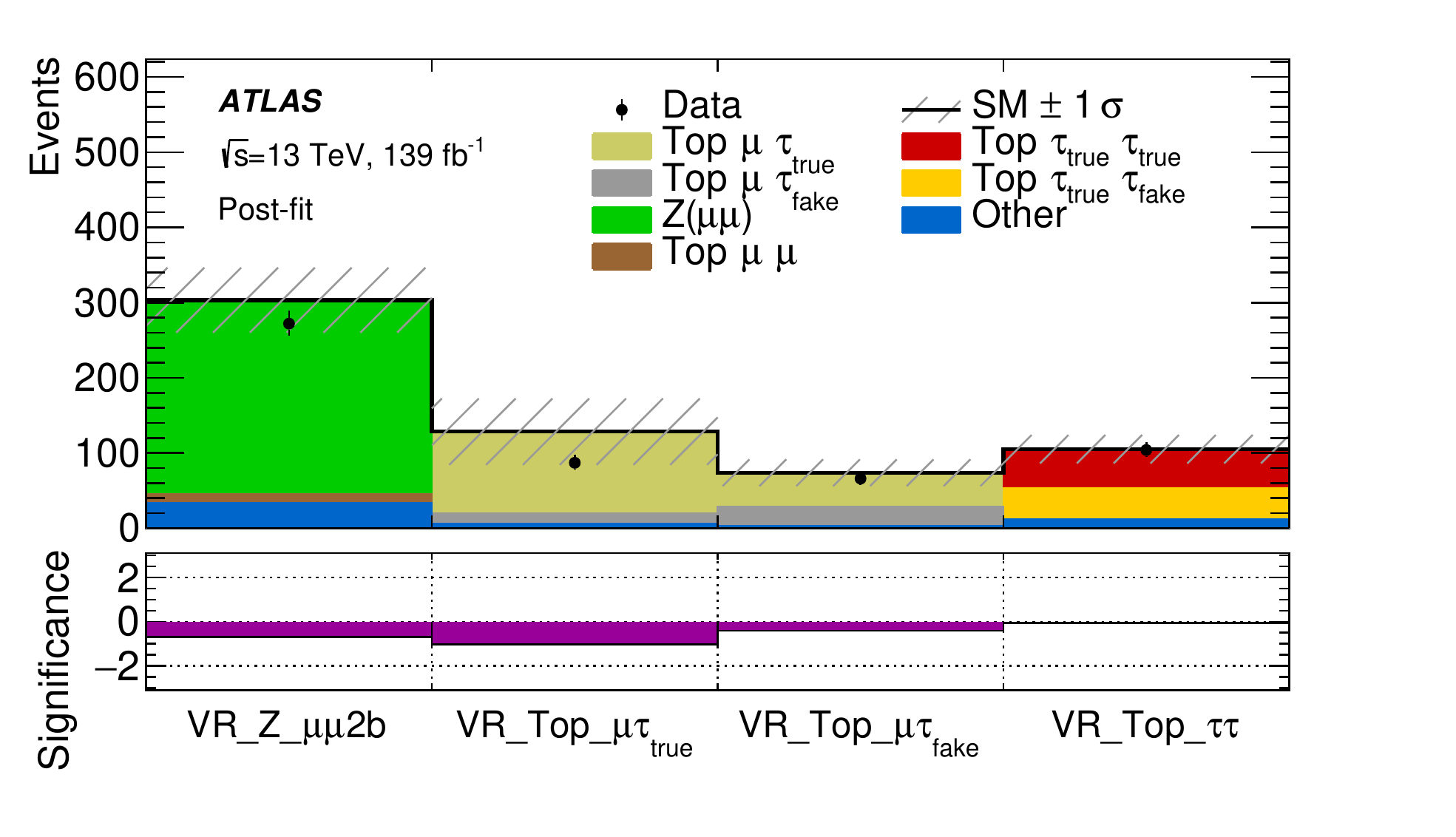}
\caption{
The upper panel shows the expected number of SM background events
and the number of events observed in data
for each of the four validation regions.
In the lower panel, the significance of the deviation of the observed yield from the expected yield is shown.
\PostfitVRRef
\BandIsStatSyst
The ``Other'' contribution includes all the backgrounds not explicitly listed in the legend (\Vjets except \Zmumujets, \ttX, diboson/triboson, multijet).
}
\label{fig:pullplots}
\end{figure}

\begin{figure}[!ht]
\centering
\subfloat[$\mintheta$ in \VRTop]{\includegraphics[width=0.3\textwidth]{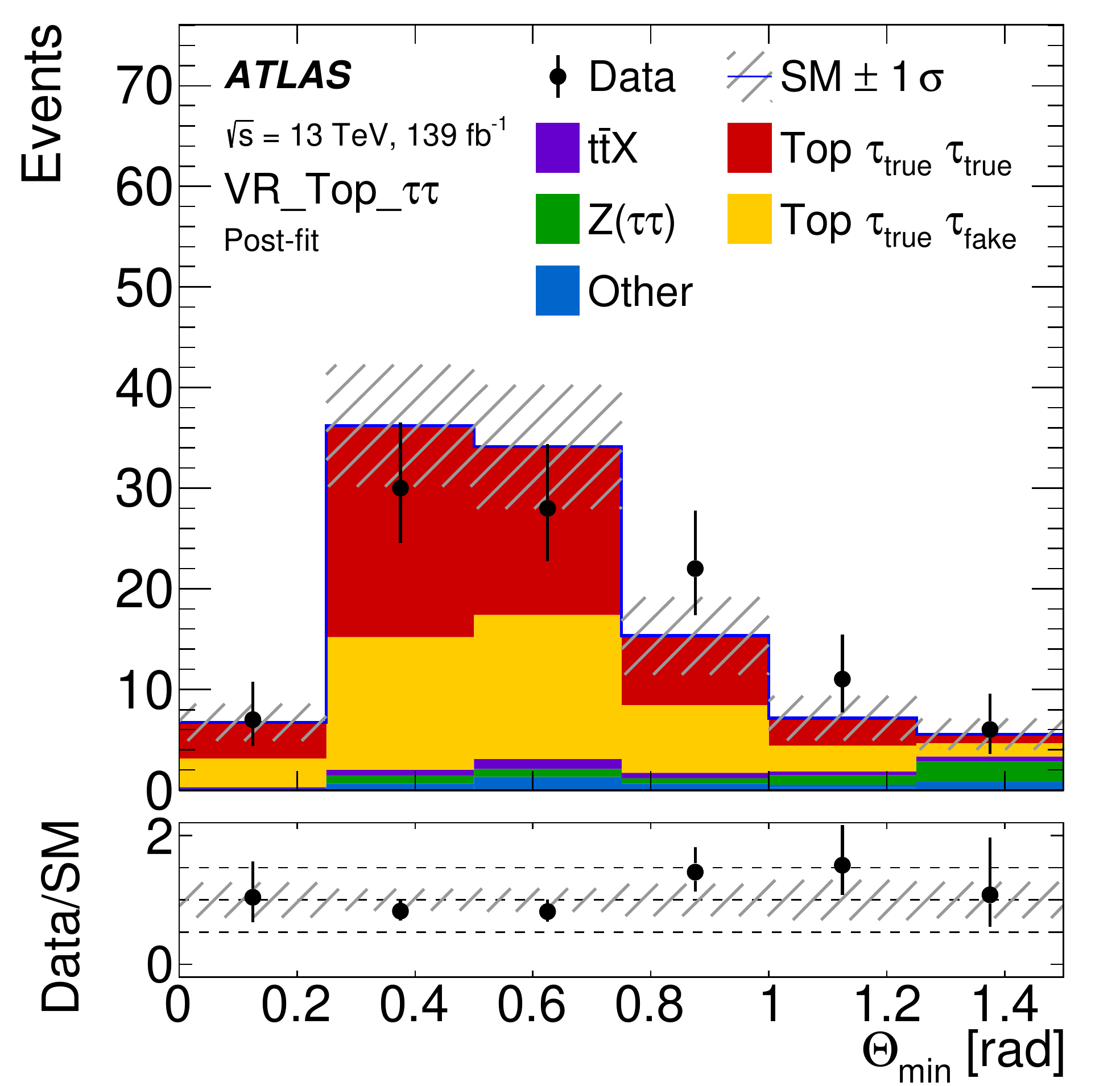} \label{fig:DitauVR_bjet_angles_afterFit}}
\hfil
\subfloat[$\mttwo$ in \VRFakeTau]{\includegraphics[width=0.3\textwidth]{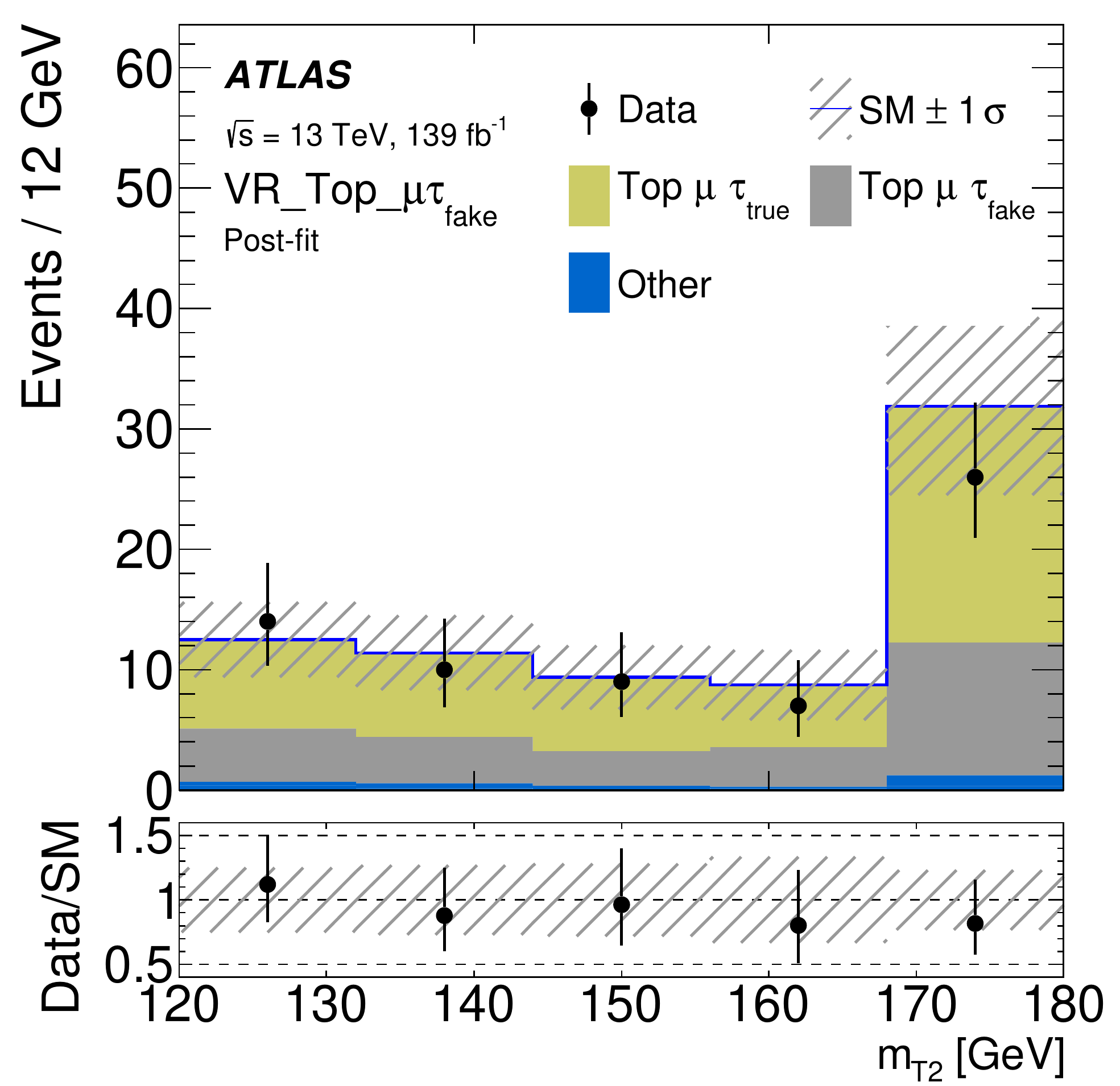} \label{fig:FakeTauVR_taujet_mt2_afterFit}}
\hfil
\subfloat[$\mttwo$ in \VRTrueTau]{\includegraphics[width=0.3\textwidth]{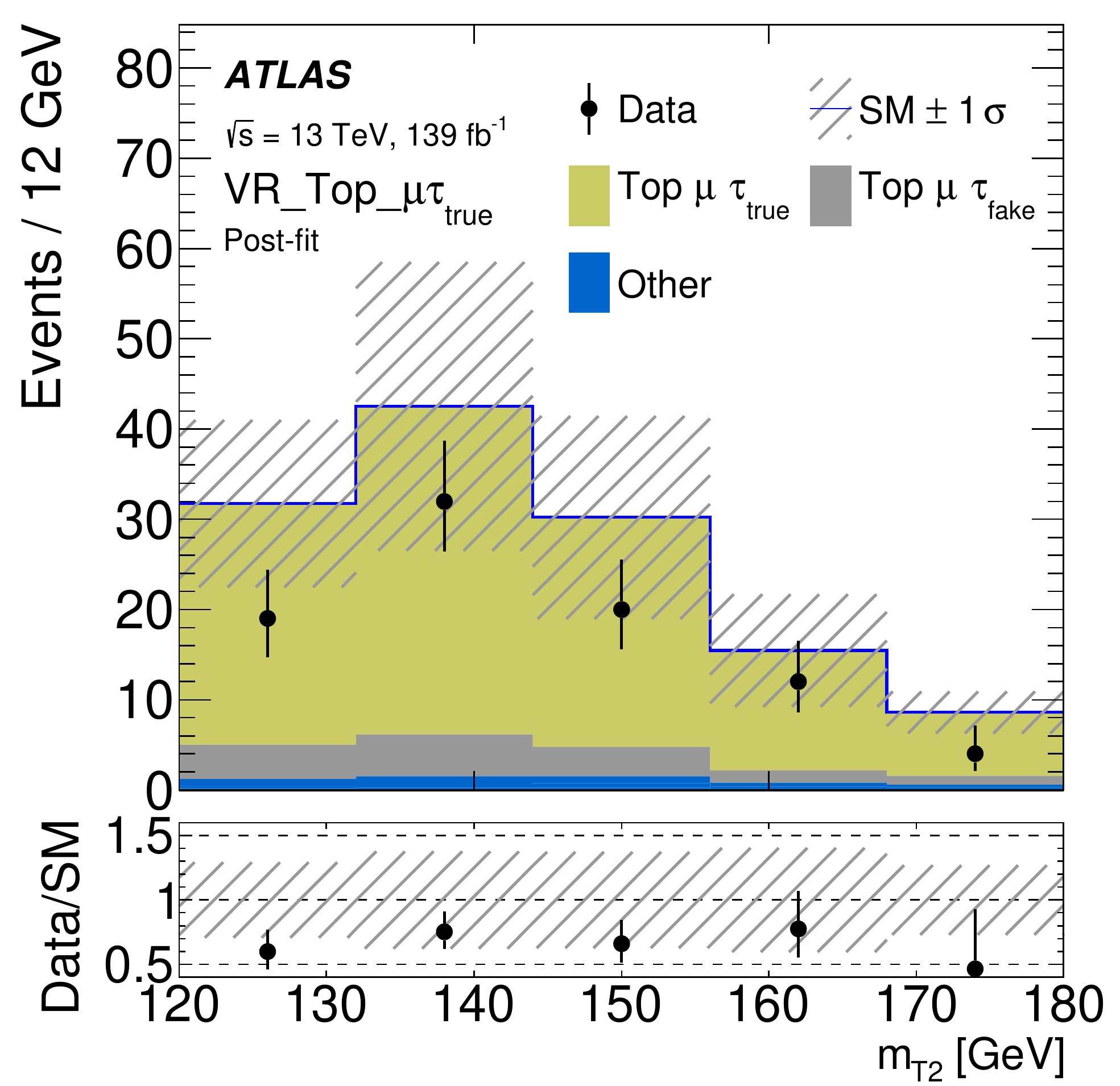}  \label{fig:TrueTauVR_taujet_mt2_afterFit}}\\
\subfloat[$\mintheta$ in \VRTrueTau]{\includegraphics[width=0.3\textwidth]{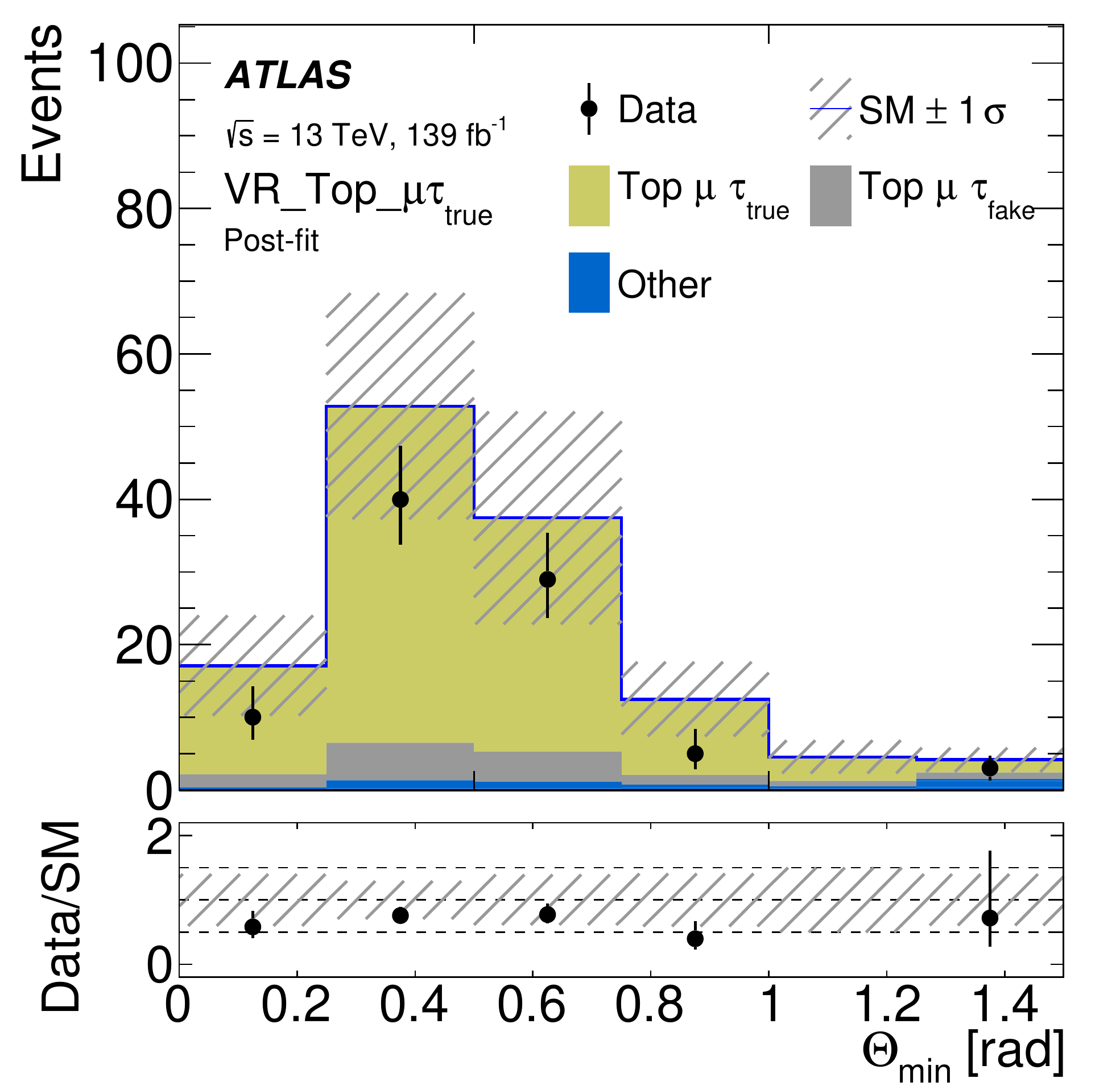} \label{fig:TrueTauVR_bjet_angles_afterFit}}
\hfil
\subfloat[$\pt(\mu_1,\mu_2)$ in \VRZmumubb]{\includegraphics[width=0.3\textwidth]{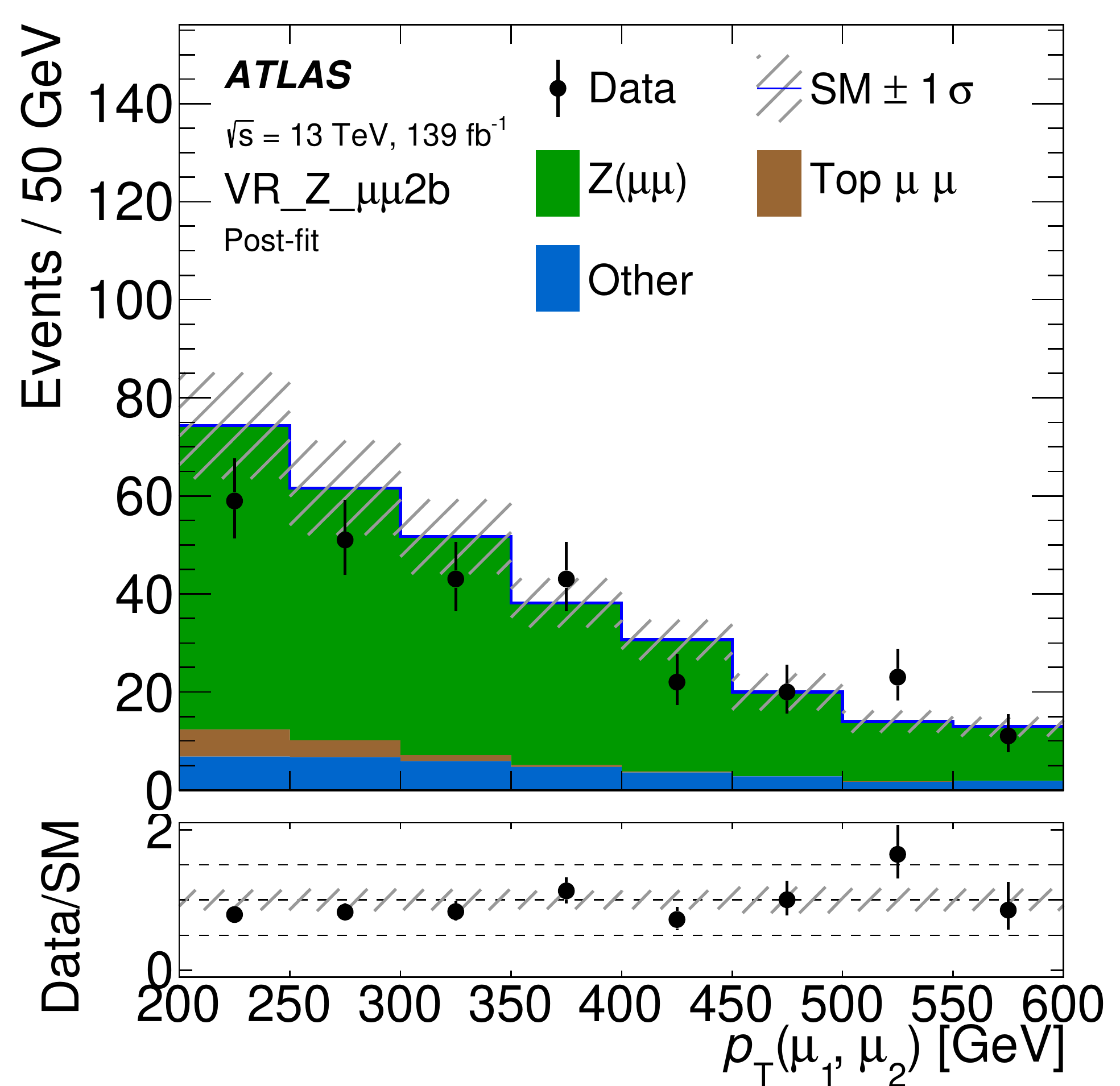} \label{fig:ZmumuVR_dimu_pt_afterFit}}
\hfil
\subfloat[$\HT$ in \VRZmumubb]{\includegraphics[width=0.3\textwidth]{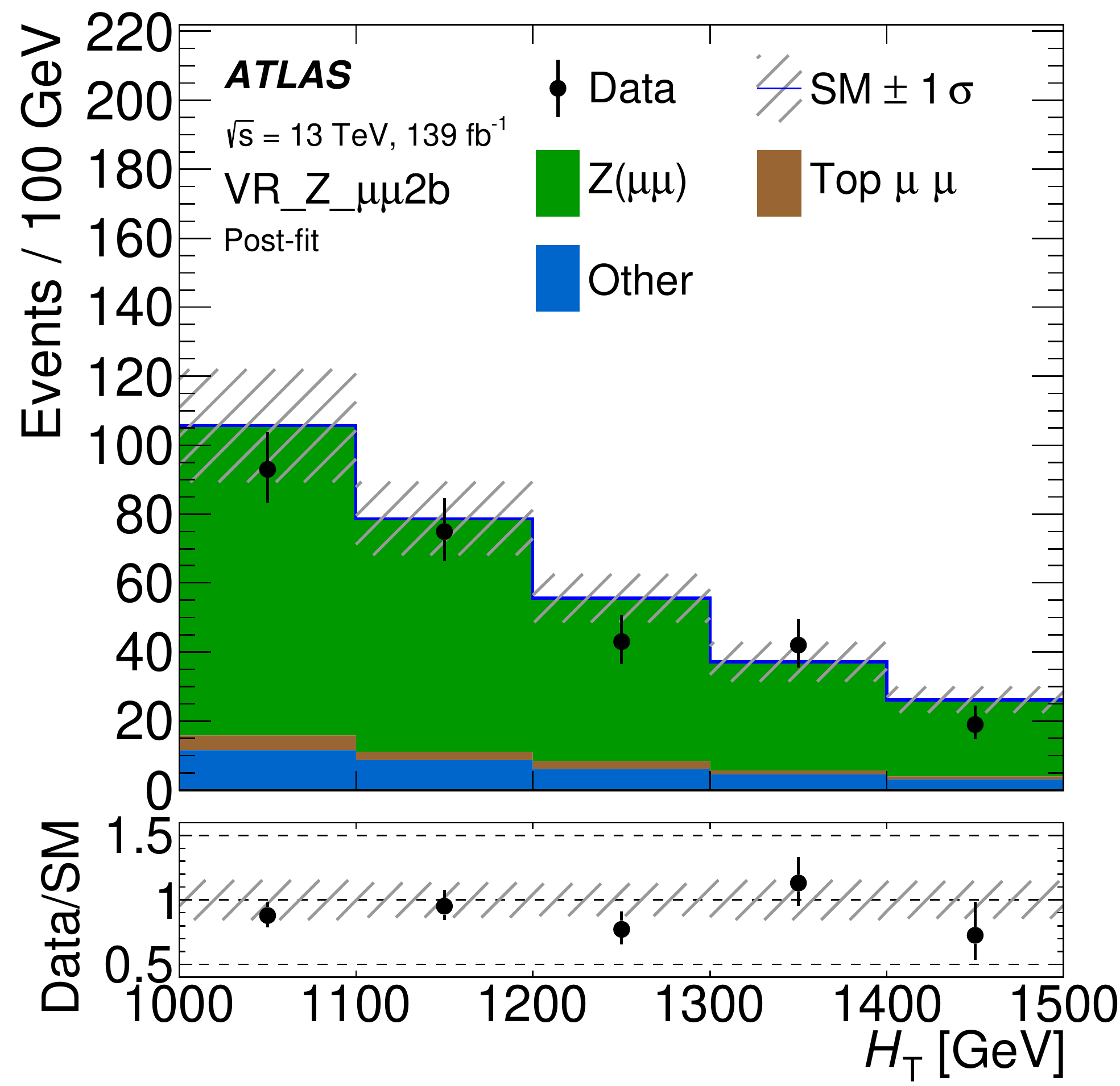}  \label{fig:ZmumuVR_ht_afterFit}}
\caption{Kinematic distributions from the four validation regions, showing
\protect\subref{fig:DitauVR_bjet_angles_afterFit} $\mintheta$ in \VRTop,
\protect\subref{fig:FakeTauVR_taujet_mt2_afterFit} $\mttwo$ in \VRFakeTau,
\protect\subref{fig:TrueTauVR_taujet_mt2_afterFit} $\mttwo$ in \VRTrueTau,
\protect\subref{fig:TrueTauVR_bjet_angles_afterFit} $\mintheta$ in \VRTrueTau,
\protect\subref{fig:ZmumuVR_dimu_pt_afterFit} $\pt(\mu_1,\mu_2)$ in \VRZmumubb,
\protect\subref{fig:ZmumuVR_ht_afterFit} $\HT$ in \VRZmumubb.
\BandIsStatSyst
\Postfit
The ``Other'' contribution includes all the backgrounds not explicitly listed in the legend (\Vjets, \ttX, diboson/triboson, multijet).
\RightBinOverflow
\BottomPanelDataSM
}
\label{fig:TopVRs}
\end{figure}

Figure~\ref{fig:pullplots} shows that the expected background yields after the fit and the observed yields
agree within one standard deviation for all four validation regions,
demonstrating good modeling of the SM background.
Figure~\ref{fig:TopVRs} shows various kinematic distributions in the validation regions.
Good agreement between the background model and the data is observed in \VRZmumubb, \VRFakeTau and \VRTop.
In \VRTrueTau, the modeling of kinematic distributions is reasonable.
The contribution of a potential signal from the model in Figure~\ref{fig:sbottom_diagram}
to the control regions does not exceed \percent{7} at the low end of the range of bottom-squark masses covered by the signal models
and quickly falls to below a percent at the high end.
For the validation regions it is around \percent{15} for low \msb and again falls to a percent or less for larger \msb.

\FloatBarrier
 
\section{Systematic uncertainties}
\label{sec:systematics}

The experimental uncertainties considered in this analysis comprise systematic uncertainties
in the reconstruction, identification, calibration and corrections applied to the physical objects used in the analysis.
They are assumed to be correlated across analysis regions and between the background processes and the signal.
Theoretical uncertainties include contributions from generator modeling as well as cross-section uncertainties.
They are assumed to be correlated across analysis regions but uncorrelated between different background processes.
When assuming no correlation between analysis regions, the total background uncertainty increases by about 5 percentage points for the single-bin SR,
and the exclusion contour does not change significantly.
 
The experimental uncertainties related to jets include uncertainties in the energy scale~\cite{PERF-2016-04} and resolution~\cite{PERF-2011-04},
jet-vertex-tagging uncertainties~\cite{ATLAS-CONF-2014-018} and flavor-tagging uncertainties~\cite{FTAG-2018-01,ATLAS-CONF-2018-001,ATLAS-CONF-2018-006}.
Flavor-related uncertainties come from the uncertainties in
data-to-simulation correction factors for efficiencies and fake rates
and from the extrapolation over jet \pt.
The $\tau$-lepton uncertainties arise from the energy calibration, and reconstruction and identification efficiencies~\cite{ATL-PHYS-PUB-2015-045,ATLAS-CONF-2017-029}.
The energy scale uncertainties include the nonclosure of the calibration and uncertainties in the detector response estimated from simulation,
as well as uncertainties in the relative calibration of data and simulation measured in $Z(\tau_\mu \tauhad)$ events.
An uncertainty at high-\pt, based on single-particle response uncertainties, is taken into account.
Muon-related uncertainties~\cite{PERF-2015-10} are not relevant in the signal regions, as events with muons do not enter these,
but they can be important in control regions with muons.
Uncertainties related to electrons have a negligible impact on this analysis.
The systematic uncertainties affecting the energy or momentum of calibrated objects are propagated to the \met calculation.
Specific uncertainties in the soft-term contribution to the \met~\cite{ATLAS-CONF-2018-023} are also considered.
 
The theoretical uncertainties related to variations of the PDFs~\cite{Butterworth:2015oua},
strong coupling constant \alphas and
renormalization and factorization scales \muR and \muF~\cite{ATL-PHYS-PUB-2018-009}
are evaluated from generator weights for all background samples.
The sets include the nominal PDF as well as $100$ variations.
The PDF uncertainty is obtained as the envelope of all the variations.
The uncertainty related to \alphas is evaluated by computing $\alphas = 0.119$ and $\alphas = 0.117$ parameterizations
and averaging the difference between them.
The PDF and \alphas uncertainties are then added in quadrature.
In order to derive the scale uncertainties, \muR and \muF are varied up and down by a factor of two.
Three independent nuisance parameters are used,
two resulting from keeping one of the scales constant while varying the other one,
and the third being the coherent variation of both scales.
The variations are normalized to the nominal sum of weights
so that the effect on the normalization included in the cross-section uncertainty is not double-counted.
For all simulated processes that are not normalized to the data,
uncertainties in the cross section and in the integrated luminosity of the data are applied.
 
For \ttbar and single-top-quark production,
generator uncertainties related to hard scattering and matching are evaluated
by comparing $\POWHEGBOX+$\pythia with \mgpythia.
Parton-showering uncertainties are estimated by comparison with $\POWHEGBOX+$\herwigseven.
Uncertainties in the initial-state and final-state radiation are evaluated by simultaneously testing the impact of scale variations
and eigenvariations of the A14 tune~\cite{ATL-PHYS-PUB-2014-021}.
For \ttbar production, an additional comparison with the \hdamp parameter set to $3 m_\text{top}$
is included.
For single-top-quark production,
an uncertainty in the treatment of the $Wt/\ttbar$ interference is considered
by comparing samples produced with the nominal diagram-removal scheme~\cite{Frixione:2008yi}
with alternative samples generated with a diagram-subtraction scheme~\cite{Frixione:2008yi,ATL-PHYS-PUB-2016-020}.
 
For the \Vjets processes,
additional uncertainties related to the resummation and CKKW matching scales~\cite{Catani:2001cc,Hoeche:2009rj}
are considered.
For the \Zmumujets and \Ztautaujets backgrounds,
the nominal \sherpa samples are compared with alternative samples produced with \mgpythia.
For diboson and \ttX samples, the PDF, scale and cross-section uncertainties are used.
 
For the bottom-squark signal samples,
uncertainties in the acceptance related to the factorization and renormalization scales, merging scales,
parton shower tuning and radiation uncertainties
are considered.
An additional uncertainty accounts for differences between samples produced with the full detector simulation and the parameterized calorimeter response.

\begin{table}[t]
\caption{
Dominant systematic uncertainties in the background prediction for the signal regions after the fit to the control regions.
Generator modeling uncertainties refer to all theoretical uncertainties, and are largely dominated by the comparisons of MC event generators for top-quark processes.
``Other'' includes the uncertainties
arising from muons, jet-vertex tagging, modeling of pileup,
the \met computation, multijet background, and luminosity.
The individual uncertainties can be correlated and do not necessarily add in quadrature to equal the total uncertainty.
}
\label{tab:SRsyst}
\centering
\begin{tabular}{lcccc}
\toprule
Uncertainty                    & Single-bin SR & \multicolumn{3}{c}{Multi-bin SR}                                \\
&               & $\mintheta < 0.5$ & $0.5 < \mintheta < 1.0$ & $\mintheta > 1.0$ \\
\midrule
Generator modeling               & $37\%$~~~        & $42\%$~~~            & $44\%$~~~                  & $27\%$~~~            \\
Normalization / transfer factors & $15\%$~~~        & $11\%$~~~            & $12\%$~~~                  & $18\%$~~~            \\
JER and JES                      & $12\%$~~~        & ~~$5.1\%$           & ~~$9.8\%$                 & $22\%$~~~            \\
$\tau$-leptons                      & ~~$8.3\%$       & ~~$3.5\%$           & ~~$2.3\%$                 & $15\%$~~~            \\
MC statistical uncertainty       & ~~$6.9\%$       & ~~$6.8\%$           & ~~$7.2\%$                 & $11\%$~~~            \\
Flavor tagging                   & ~~$3.8\%$       & ~~$1.0\%$           & ~~$1.8\%$                 & ~~$5.4\%$           \\
Other                            & ~~$2.9\%$       & ~~$1.3\%$           & ~~$1.8\%$                 & ~~$6.6\%$           \\
\midrule
Total                            & $40\%$~~~        & $43\%$~~~            & $46\%$~~~                  & $41\%$~~~            \\
\bottomrule
\end{tabular}
\end{table}

A summary of the dominant systematic uncertainties in the background prediction for the signal regions is given in Table~\ref{tab:SRsyst}.
The largest source of uncertainty is the generator modeling,
and here in particular the modeling of the top-quark background,
mainly the modeling of the hard-scatter process and initial state radiation uncertainties.
Second-leading in size is the total uncertainty in the normalization and transfer factors,
which is obtained from the fit.
As the transfer factors are ratios of normalization factors,
and a large part of the uncertainties cancel out in the ratio,
the uncertainties in the transfer factors are comparatively small.

 
\section{Results}
\label{sec:result}

\begin{table}[t]
\renewcommand{\arraystretch}{1.1}
\caption{
The observed event yields in data,
the total expected yields from \SM processes obtained from the background-only fit and breakdown of individual contributions,
and the expected signal contributions for three benchmark models
are shown for the single-bin signal region and the three bins of the multi-bin signal region.
Total uncertainties combining the statistical and systematic uncertainties are quoted for the background processes.
For the signal, the quoted uncertainties are only statistical.
``Other'' combines all \SM background contributions that are not listed explicitly,
covering \Vjets except for \Ztautaujets, multijet, diboson and triboson contributions.
The dash means that no events pass the selection.
}
\label{tab:sr1yield}
\centering
\begin{tabular}{lcccc}
\toprule
& Single-bin SR               & \multicolumn{3}{c}{Multi-bin SR}                                                          \\
&                             & $\mintheta < 0.5$           & $0.5 < \mintheta < 1.0$       & $\mintheta > 1.0$           \\
\midrule
Observed events                         & $4$                         & $3$                         & $1$                           & $3$                         \\
\midrule
Total SM background                     & $3.8 \pm 1.5$               & $2.7 \pm 1.1$               & $3.5 \pm 1.6$                 & $1.5 \pm 0.6$               \\
\midrule
~~Top quark \ttrue\ttrue                & $1.4 \pm 0.9$               & $1.6 \pm 0.7$               & $1.9 \pm 1.0$                 & $0.30~_{-~0.30}^{+~0.41}$~~ \\
~~Top quark \ttrue\tfake                & $0.92 \pm 0.62$             & $0.76 \pm 0.43$             & $0.96 \pm 0.69$               & $0.22 \pm 0.17$             \\
~~Top quark \tfake\tfake                & $0.11~_{-~0.11}^{+~0.26}$~~ & $0.06 \pm 0.06$             & $0.12~_{-~0.12}^{+~0.23}$~~   & $0.04~_{-~0.04}^{+~0.05}$~~ \\
~~\ttX                                  & $0.52 \pm 0.42$             & $0.18 \pm 0.10$             & $0.26~_{-~0.26}^{+~0.31}$~~   & $0.31 \pm 0.22$             \\
~~\Ztautaujets                          & $0.73 \pm 0.25$             & $0.05 \pm 0.05$             & $0.17 \pm 0.16$               & $0.59 \pm 0.22$             \\
~~Other                                 & $0.07 \pm 0.04$             & ---                         & $0.04 \pm 0.01$               & $0.06 \pm 0.03$             \\
\midrule
$m(\sbottom,\ninotwo) = (800,131)$~\gev & $5.6  \pm 1.4$              & $0.14 \pm 0.06$             & $1.5 \pm 0.4$                 & $4.3 \pm 1.1$               \\
$m(\sbottom,\ninotwo) = (800,180)$~\gev & $9.3  \pm 2.2$              & $0.08~_{-~0.08}^{+~0.14}$~~ & $2.4 \pm 0.6$                 & $7.1 \pm 1.7$               \\
$m(\sbottom,\ninotwo) = (350,280)$~\gev & $6.4  \pm 2.1$              & $2.7 \pm 0.9$               & $4.1 \pm 1.3$                 & $4.8 \pm 1.8$               \\
\bottomrule
\end{tabular}
\end{table}

The event yields for all signal regions are reported in Table~\ref{tab:sr1yield}.
The \SM background prediction is based on the background-only fit described in Section~\ref{sec:background}.
To illustrate the order of magnitude of the contribution of signal events, the expected yields for three benchmark signal models are included in the table.
The single-bin SR and the first two bins of the multi-bin SR are dominated by top-quark production,
whereas for $\mintheta > 1.0$ the \ztautau background is the largest
contribution.
Other \SM processes contribute very little to the signal regions.
Figure~\ref{fig:SRyields} shows a comparison of data and background yields in the SRs
together with the corresponding significances
quantifying the deviation of the observed yields from the \SM expectation
in the bottom panel.
No significant excess of data above the expected yields from the \SM background processes is observed in any of the signal regions.
The $p$-value for the event yield in the single-bin signal region to fluctuate to at least the observed value
under the background-only hypothesis is $p(s=0) = 0.44$.

\begin{figure}[t]
\centering
\includegraphics[width=0.9\textwidth]{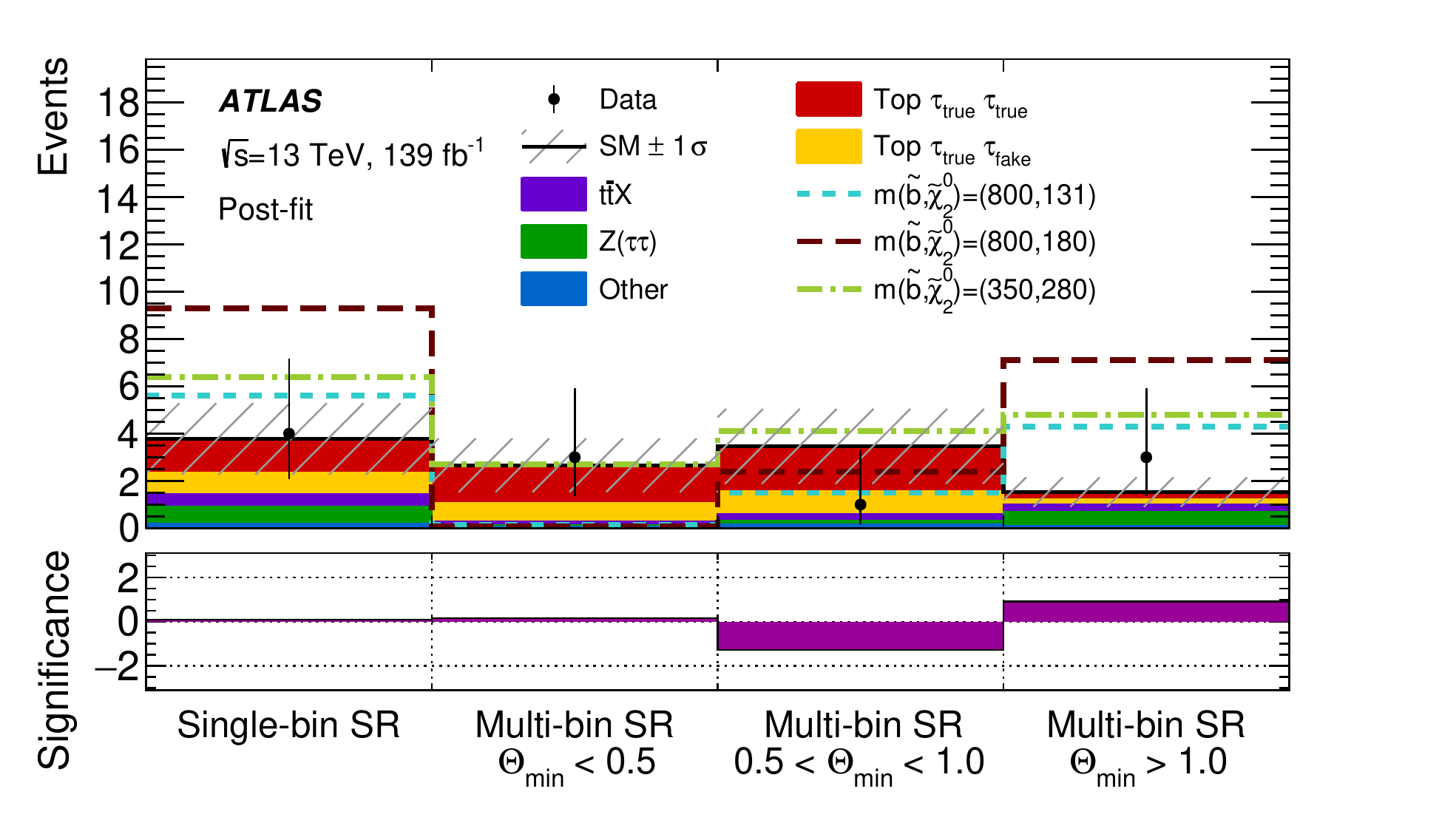}
\caption{
Comparison of the expected and observed event yields in the signal regions defined in~\cref{tab:SR_cuts}.
\Postfit
The ``Other'' contribution includes all the backgrounds not explicitly listed in the legend (\Vjets except \Ztautaujets, diboson/triboson, multijet).
\BandIsStatSyst
The contributions from three signal models to the signal regions are also displayed,
where the masses \msb and \mneuttwo are given in \gev\ in the legend.
The lower panel shows
the significance of the deviation of the observed yield from the expected background yield.
}
\label{fig:SRyields}
\end{figure}

\begin{figure}[htb]
\centering
\includegraphics[width=1.0\textwidth]{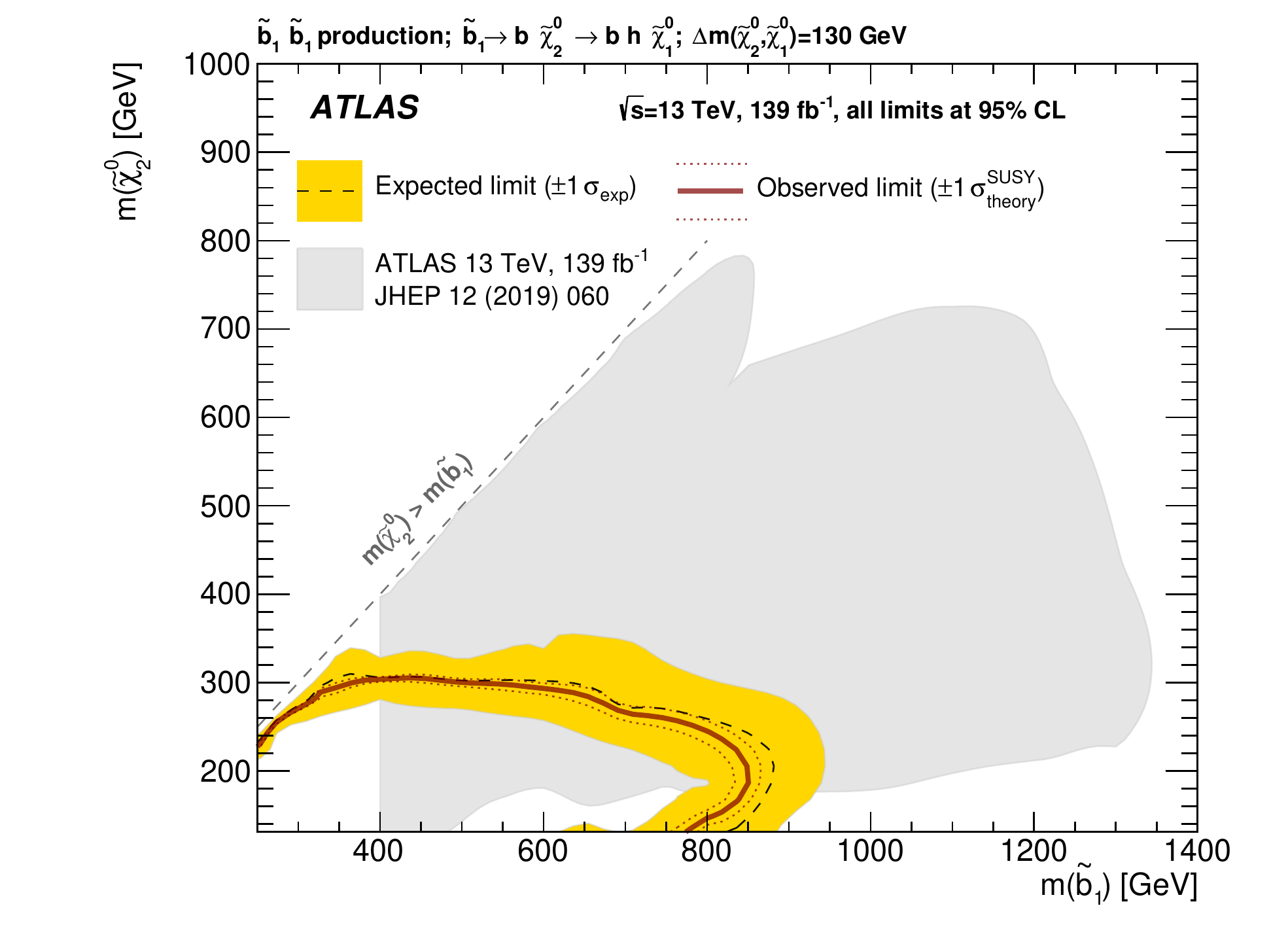}
\caption{
Exclusion contours at the \percent{95} CL as a function of \msb and \mneuttwo,
assuming $\Delta m(\ninotwo, \ninoone) = \GeVx{130}$.
Observed and expected limits are shown for the present search that requires hadronically decaying $\tau$-leptons, \bjets and \met in the final state.
The observed exclusion limit from a previous ATLAS search~\cite{SUSY-2018-31} that requires \bjets and \met in the final state is also displayed.
The region $m(\sbottom) < \GeVx{400}$ is excluded by a previous search from CMS~\cite{CMS-SUS-18-007}.
}
\label{fig:exclusion_combination}
\end{figure}

Exclusion contours at the \percent{95} confidence level (CL) are derived from the yields in the multi-bin signal region
for the two-dimensional parameter space of \msb and \mneuttwo in the simplified model from Figure~\ref{fig:sbottom_diagram}.
A fixed mass difference of \GeVx{130} between the second-lightest neutralino \ninotwo and lightest neutralino \ninoone is assumed for all signal models.
The probabilities that the data are compatible with the background-only and signal-plus-background hypotheses are evaluated
using a one-sided profile-likelihood-ratio test statistic and the \CLs prescription~\cite{Read:2002hq}.
The computations rely on asymptotic properties of the profile-likelihood ratio~\cite{Cowan:2010js}.
Systematic uncertainties are treated as nuisance parameters with Gaussian probability densities in the likelihood function.
The resulting observed and expected exclusion contours are shown in Figure~\ref{fig:exclusion_combination}.
The uncertainties in the cross section of the supersymmetric signal are not included in the fit
but shown as an uncertainty band around the observed limit contour.
Since the observed data yield is larger than the expected total background in the highest \mintheta bin,
which is most sensitive to models with large \msb,
the observed exclusion contour deviates inwards from the expected contour with increasing \msb,
but it stays within the uncertainty band of the expected limit.
The search is optimized for the low-\mneuttwo region and has sensitivity to models with \mneuttwo up to \GeVx{300}.
Bottom squarks with masses up to \GeVx{850} are excluded in this region.
For \mneuttwo below about \GeVx{200}, the softer \met spectrum of the signal results in a lower acceptance,
leading to a slightly reduced exclusion reach in bottom-squark mass.
The parameter-space region where $\Delta m(\sbottom,\ninotwo) \lesssim \GeVx{20}$ cannot be excluded as the bottom-squark decay products are not boosted enough,
and the stringent kinematic requirements in the SRs result in low signal acceptance.
These results are overlaid on the observed exclusion contour from a previous ATLAS search~\cite{SUSY-2018-31}
to demonstrate the complementarity of the two approaches.
The new results have unique sensitivity to a previously uncovered region of parameter space at low \ninotwo masses,
where the previous search quickly loses sensitivity.

The results from the single-bin signal region can be interpreted in terms of model-independent upper limits on the event yields
from potential BSM processes.
The fit is performed simultaneously over the CRs and the single-bin SR, assuming no signal contribution in the CRs.
The profile-likelihood-ratio test statistic is evaluated using pseudo-experiments.
An upper limit of \fbx{0.05} is derived for the visible cross section \sigvis,
defined as the product of the cross section, acceptance and selection efficiency of such processes.
In addition, Table~\ref{tab:discSR} summarizes the expected and observed \percent{95} CL upper limits on the number of BSM events,
as well as the confidence level of the background-only hypothesis \CLb.
The $p$-value and the corresponding significance for the background-only hypothesis to fluctuate to at least the observed values are also included.
 
\begin{table}[t]
\caption{
Upper limits at \percent{95} CL on the visible cross section \sigvis,
on the number of signal events ($S_\mathrm{obs}^{95}$),
and on number of signal events given the expected number (and $\pm 1\sigma$ excursions of the expectation) of background events ($S_\mathrm{exp}^{95}$).
The last two columns indicate the \CLb value,
\ie the confidence level observed for the background-only hypothesis,
the discovery $p$-value ($p(s = 0)$),
and its associated significance $Z$.
}
\label{tab:discSR}
\centering
\begin{tabular}{lccccc}
\toprule{Signal region} & \sigvis[fb]  &  $S_\mathrm{obs}^{95}$  & $S_\mathrm{exp}^{95}$ & \CLb & $p(s=0)$ ($Z$)  \\
\midrule
Single-bin SR & $0.05$ &  $6.6$ & $ { 6.0 }^{ +2.3 }_{ -1.6 }$ & $0.62$ & $ 0.34$~$(0.41)$ \\
\bottomrule
\end{tabular}
\end{table}

\FloatBarrier
 
\section{Conclusion}
\label{sec:conclusion}
 
A search for bottom-squark pairs in events with \bjets, hadronically decaying $\tau$-leptons and large missing transverse momentum is presented.
A simplified SUSY model assuming $\sbottom \to b \ninotwo \to b h \ninoone$ is considered, where at least one Higgs boson decays into a pair of $\tau$-leptons.
This analysis has unique sensitivity at low \ninotwo masses due to the presence of hadronically decaying $\tau$-leptons,
which mitigates the Standard Model background,
and to the associated $\nu_\tau$-neutrinos that add to the \met originating from the \ninoone.
A multi-bin signal region exploiting angular correlations between the \bjets and the hadronically decaying $\tau$-leptons is used to search for a \sbottom signal,
and a single-bin signal region is employed for a model-independent statistical interpretation.
The data observed in the signal regions are compatible with the expected Standard Model background.
Exclusion limits are placed on the bottom-squark mass at the \percent{95} confidence level.
For \mneuttwo ranging from \GeVx{130} to \GeVx{180},
bottom-squark masses below \GeVx{775} to \GeVx{850} are excluded.
This extends significantly beyond the reach of a previous ATLAS search~\cite{SUSY-2018-31},
which was performed in final states with \bjets and large \met,
in this challenging region of parameter space.
 
\section*{Acknowledgments}

We thank CERN for the very successful operation of the LHC, as well as the
support staff from our institutions without whom ATLAS could not be
operated efficiently.
 
We acknowledge the support of
ANPCyT, Argentina;
YerPhI, Armenia;
ARC, Australia;
BMWFW and FWF, Austria;
ANAS, Azerbaijan;
SSTC, Belarus;
CNPq and FAPESP, Brazil;
NSERC, NRC and CFI, Canada;
CERN;
ANID, Chile;
CAS, MOST and NSFC, China;
Minciencias, Colombia;
MEYS CR, Czech Republic;
DNRF and DNSRC, Denmark;
IN2P3-CNRS and CEA-DRF/IRFU, France;
SRNSFG, Georgia;
BMBF, HGF and MPG, Germany;
GSRI, Greece;
RGC and Hong Kong SAR, China;
ISF and Benoziyo Center, Israel;
INFN, Italy;
MEXT and JSPS, Japan;
CNRST, Morocco;
NWO, Netherlands;
RCN, Norway;
MEiN, Poland;
FCT, Portugal;
MNE/IFA, Romania;
JINR;
MES of Russia and NRC KI, Russian Federation;
MESTD, Serbia;
MSSR, Slovakia;
ARRS and MIZ\v{S}, Slovenia;
DSI/NRF, South Africa;
MICINN, Spain;
SRC and Wallenberg Foundation, Sweden;
SERI, SNSF and Cantons of Bern and Geneva, Switzerland;
MOST, Taiwan;
TAEK, Turkey;
STFC, United Kingdom;
DOE and NSF, United States of America.
In addition, individual groups and members have received support from
BCKDF, CANARIE, Compute Canada and CRC, Canada;
COST, ERC, ERDF, Horizon 2020 and Marie Sk{\l}odowska-Curie Actions, European Union;
Investissements d'Avenir Labex, Investissements d'Avenir Idex and ANR, France;
DFG and AvH Foundation, Germany;
Herakleitos, Thales and Aristeia programmes co-financed by EU-ESF and the Greek NSRF, Greece;
BSF-NSF and GIF, Israel;
Norwegian Financial Mechanism 2014-2021, Norway;
NCN and NAWA, Poland;
La Caixa Banking Foundation, CERCA Programme Generalitat de Catalunya and PROMETEO and GenT Programmes Generalitat Valenciana, Spain;
G\"{o}ran Gustafssons Stiftelse, Sweden;
The Royal Society and Leverhulme Trust, United Kingdom.
 
The crucial computing support from all WLCG partners is acknowledged gratefully, in particular from CERN, the ATLAS Tier-1 facilities at TRIUMF (Canada), NDGF (Denmark, Norway, Sweden), CC-IN2P3 (France), KIT/GridKA (Germany), INFN-CNAF (Italy), NL-T1 (Netherlands), PIC (Spain), ASGC (Taiwan), RAL (UK) and BNL (USA), the Tier-2 facilities worldwide and large non-WLCG resource providers. Major contributors of computing resources are listed in Ref.~\cite{ATL-SOFT-PUB-2021-003}.
 

\clearpage
\printbibliography
 
\clearpage
 
\begin{flushleft}
\hypersetup{urlcolor=black}
{\Large The ATLAS Collaboration}

\bigskip

\AtlasOrcid[0000-0002-6665-4934]{G.~Aad}$^\textrm{\scriptsize 102}$,    
\AtlasOrcid[0000-0002-5888-2734]{B.~Abbott}$^\textrm{\scriptsize 128}$,    
\AtlasOrcid[0000-0002-7248-3203]{D.C.~Abbott}$^\textrm{\scriptsize 103}$,    
\AtlasOrcid[0000-0002-2788-3822]{A.~Abed~Abud}$^\textrm{\scriptsize 36}$,    
\AtlasOrcid[0000-0002-1002-1652]{K.~Abeling}$^\textrm{\scriptsize 53}$,    
\AtlasOrcid[0000-0002-2987-4006]{D.K.~Abhayasinghe}$^\textrm{\scriptsize 94}$,    
\AtlasOrcid[0000-0002-8496-9294]{S.H.~Abidi}$^\textrm{\scriptsize 167}$,    
\AtlasOrcid[0000-0002-8279-9324]{O.S.~AbouZeid}$^\textrm{\scriptsize 40}$,    
\AtlasOrcid{N.L.~Abraham}$^\textrm{\scriptsize 156}$,    
\AtlasOrcid[0000-0001-5329-6640]{H.~Abramowicz}$^\textrm{\scriptsize 161}$
% ...
\bigskip
\\

% ...
$^{36}$CERN, Geneva; Switzerland.\\
% ...
$^{40}$Niels Bohr Institute, University of Copenhagen, Copenhagen; Denmark.\\
% ...
$^{53}$II. Physikalisches Institut, Georg-August-Universit\"{a}t G\"ottingen, G\"ottingen; Germany.\\
% ...
$^{94}$Department of Physics, Royal Holloway University of London, Egham; United Kingdom.\\
% ...
$^{102}$CPPM, Aix-Marseille Universit\'e, CNRS/IN2P3, Marseille; France.\\
$^{103}$Department of Physics, University of Massachusetts, Amherst MA; United States of America.\\
% ...
$^{128}$Homer L. Dodge Department of Physics and Astronomy, University of Oklahoma, Norman OK; United States of America.\\
% ...
$^{156}$Department of Physics and Astronomy, University of Sussex, Brighton; United Kingdom.\\
% ...
$^{161}$Raymond and Beverly Sackler School of Physics and Astronomy, Tel Aviv University, Tel Aviv; Israel.\\
% ...
$^{167}$Department of Physics, University of Toronto, Toronto ON; Canada.\\
% ...

\end{flushleft}


\end{document}